\documentclass[11pt]{article}
	\usepackage[utf8]{inputenc}
	\usepackage{blindtext}
	\usepackage[utf8]{inputenc}
	\usepackage{times}
	\usepackage[left=1.8cm,top=2.2cm,right=1.8cm,bottom=2.2cm]{geometry}
	\usepackage{inputenc}
	\usepackage{multicol}
	\usepackage{multirow}
	\usepackage{graphicx}
	\usepackage{booktabs} 
	\usepackage{array}
	\usepackage{paralist}
	\usepackage{verbatim} 
	\usepackage{subfigure}
	\usepackage{subfig}
	\usepackage[usenames]{color}	
	\usepackage{fixltx2e}
	\usepackage{float}
	\usepackage{ulem}
	\usepackage{fancyhdr}
\title{\textbf{Aproximación a la estimación espacial de sequías meteorológicas en la cuenca hidrográfica del río Coello, Colombia}}

\author{Cruz Roa Andrés Felipe$^{1}$\footnote{Trabajo de grado para optar el titulo de Ingeniero Forestal, Universidad del Tolima.} \& Barrios Peña Miguel Ignacio$^{2}$\\
\small{$^{1}$Facultad de Ingeniería Forestal,  $^{2}$Departamento de Ingeniería de la Facultad de Ingeniería Forestal}\\
\small{Universidad del Tolima, 730006299, Ibagué, Colombia.} \small{\texttt{\ $^{1}$afcruz@ut.edu.co, $^{2}$mibarrios@ut.edu.co}}}
\date{\small{\today}}

\begin{document}

\maketitle

\small{\textbf{Resumen:}
La evaluación espacio temporal para la caracterización de las sequías meteorológicas se basó en datos de precipitación mensual acumulada entre 1996-2005 de 20 estaciones meteorológicas distribuidas en la cuenca hidrográfica del río Coello. Se realizó un preprocesamiento a los datos de precipitación con pruebas de consistencia de datos para corregir y eliminar datos sobre o sub estimados. Para estimar los datos faltantes de precipitación se comparan tres métodos geoestadísticos de interpolación, derivados de Kriging, asociados con variables secundarias como el Kriging Ordinario, CoKrigin Ordinario asocidadas a las variables secundarias de un Modelo de Elevación Digital y datos de precipitación satelital TRMM. Para seleccionar el método geoestadístico se comparó el ajuste de cada interpolación con respecto a tres estaciones de referencia a través de tres pruebas de calidad, las cuales fueron Raíz del Error Cuadrático Medio (RMSE), Criterio de Información de Akaike (AIC) y Criterio de Información Bayesiano (BIC). En esta investigación dos de tres pruebas favorecen al CoKriging Ordinario usando como variable secundaria la Altitud (CoK+DEM). Con la serie interpolada de precipitación se evaluaron y caracterizaron las sequías por medio del Índice de Precipitación Estandarizado (SPI) a escala mensual y trimestral, calculando los parámetros de severidad, duración, intensidad y frecuencia de las sequías. Por medio de mapas se delimitaron las regiones en donde se presentan los valores negativos de SPI. En el análisis espacio temporal los meses de Enero, Febrero, Julio y Agosto son los más secos del año. En el año 1997 se presentó la sequía meteorológica de mayores afectaciones en la cuenca del río Coello generalmente concentrados en la parte media y baja de la cuenca, con una intensidad maxima de -2,57 de SPI.
\smallskip

\textbf{Palabras Clave:} Sequía meteorológica; métodos de interpolación geoestadísticos; Índice de Precipitación Estandarizado (SPI); parámetros de sequías; cuenca hidrográfica del río Coello.

\medskip
\begin{center}
\textbf{Approach to spatial estimation of meteorological droughts in the Coello River basin, Colombia}
\end{center}

\textbf{Abstract:}
The space-temporal evaluation to characterize meteorological droughts was based on data accumulated monthly precipitation between 1996-2005 from 20 meteorological stations distributed in the Coello River basin. Data precipitation was performed preprocessing with data consistency tests to correct and delete data over- or under estimated. To estimate missing precipitation data are compared three geostatistical interpolation methods derived from Kriging, associated with secondary variables such as the Ordinary Kriging, CoKrigin Ordinary associated with secondary variables of a Digital Elevation Model and data satellite TRMM. To select the statistical method the setting of each interpolation was compared with respect to three reference stations through three quality tests, which were Root Mean Square Error (RMSE), Akaike Information Criterion (AIC) and Bayesian Information Criterion (BIC). In this investigation two of three tests favor the Ordinary CoKriging using as a secondary variable Altitude (COK+DEM). With the interpolated series of precipitation were evaluated and characterized by drought Standardized Precipitation Index (SPI) at monthly and quarterly scale, calculating the parameters of severity, duration, intensity and frequency of droughts. By mapping are delimited the regions where occur the more negative values of SPI. In analyzing space-temporal the months of January, February, July and August are the driest of the year. In 1997 the meteorological drought greatest damage occurs in the Coello River basin generally concentrated in the middle and lower part of the basin, with a maximum intensity of SPI -2,57.
\smallskip



\begin{multicols}{2}
	\section{\normalsize{INTRODUCCIÓN}}
Las sequías meteorológicas son fenómenos que se presentan por déficit de precipitación en un período continuo, disminuyendo la humedad relativa ambiental, aumentando de la temperatura y reduciendo la tasa de recarga de aguas subterráneas \textcolor{blue}{(Mishra \& Singh, 2010; Palmer, 1965; Valiente, 2001)}. Para entender las relaciones entre las sequías meteorológicas y los patrones climáticos es necesario realizar investigaciones cuantitativas en áreas de importancia para disminuir y mitigar los impactos negativos de estos fenómenos, sirviendo como una herramienta de planificación y manejo integral de los recursos hídricos \textcolor{blue}{(Mendoza \& Puche, 2003; Mishra \& Singh, 2011; Valdés $et$ $al$, 2008)}.
\smallskip

El Índice de Precipitación Estandarizado (SPI) desarrollado por \textcolor{blue}{Mckee $et$ $al$ (1993)} es el índice más utilizado en el campo de la hidrología por su simplicidad y versatilidad para el monitoreo de periodos secos y húmedos \textcolor{blue}{(Svoboda $et$ $al$, 2012; Quiring, 2009)}. \textcolor{blue}{McKee $et$ $al$ (1995)} realizaron con datos mensuales de precipitación, en un intervalo de dos años, en el estado de Colorado (USA), el calculo de SPI de tres, seis, doce y veinticuatro meses, describiendo por medio de mapas los periodos de sequía, validando el índice y sugiriendo escalas temporales más amplias. En los registros diarios de precipitación de las estaciones meteorológicas utilizados para el cálculo del SPI y parámetros, se presentan errores instrumentales y/o de medición. Para completar los datos erróneos o faltantes de precipitación se utilizan los Métodos de Interpolación Geoestadísticos como el Kriging Ordinario (OK); El $OK$ es un método interpolador que obtiene una estimación lineal de los valores ajustados a un variograma con la información disponible, obteniendo valores con mínima varianza de estimación. El $OK$ puede considerar variables secundarias, que sirven para correlacionarlas y estimar un valor más aproximado a la realidad, en algunos casos la Altitud y datos de precipitación satelital \textcolor{blue}{(Bayat $et$ $al$, 2014; Oliver \& Webster, 2015; Serrat-Capdevila $et$ $al$, 2014)}.

\textcolor{blue}{Bayat $et$ $al$ (2014)} realizó un estudio hidrológico, utilizando 94 estaciones meteorológicas para el calculo el SPI en la serie de tiempo 1966 al 2005, en donde comparó métodos no-geoestadísticos y geoestadisticos para estimar la severidad de sequías meteorológicas en la cuenca del lago Namak, parte central de Irán. En este estudio se establecieron las regiones más propensas a las sequías y delimitaron las áreas productivas más sensibles y resistentes, sirviendo como una herramienta de planificación para la toma de decisiones. Las investigaciones conceptuales de \textcolor{blue}{He $et$ $al$ (2014)} y \textcolor{blue}{Anderegg $et$ $al$ (2012)} manifiestan que las sequías conllevan al estrés físico a los ecosistemas boscosos que inducen a la mortalidad de árboles y que favorecen a los incendios forestales en áreas vulnerables, sumando la influencia global del cambio climático que contribuye a que estos efectos negativos sean más frecuentes y mayor magnitud en los ecosistemas terrestres. \textcolor{blue}{Gocic \& Trajkovic (2014)} realizaron una caracterización espacio temporal entre 1948 al 2012 de las sequías en Serbia, confrontando el SPI y el S-mode de los Análisis de Componentes Principales (PCA), utilizando escalas temporales para 12 y 24 meses. Para la confrontación de métodos, no se encontraron diferencias entre los dos índices utilizados. En el estudio se establecío que el 70\% de las sequías se categorizaron como normales, pero en los años 1990, 2000 y 2011 se detectaron las sequías con categorias de severas a extremadamente secas. El período comprendido entre 1961 y 1990 coinciden como climas normales. Esta investigación contribuyó a la planificación ambiental de los recursos naturales, demostrando lo importante y versátil que resulta ser el SPI para caracterizar las sequías. \textcolor{blue}{Patel $et$ $al$ (2007)} calcularon el SPI para 23 años (1981-2003) en el estado de Guajarat, India occidental. Esta investigación releva que los estudios trimestrales de SPI son los más importante para la caracterización de sequías y establece que este índice tiene una asociación directa con la producciones agrícolas de las regiones evaluadas. De igual forma \textcolor{blue}{Vicente-Serrano (2006)} investigó la evolución de las sequías en la Península Ibérica en España, desde 1910 al 2000. Evaluó las sequías con el análisis anual de SPI. La investigación establece que los años 1940, 1950, 1980 y los 1990 fueron los años donde mayor intensidad tuvieron las sequías en el área de estudio y la distribución espacial fue diversa. Por último concluye que la demanda hídrica del sector urbano influye en la dinámica de sequías. \textcolor{blue}{Sirdaş \& Sen (2003)} utilizarón el OK para realizar los análisis espaciales de sequías, produciendo mapas de SPI para la región de Trakya, en Turquía, entre los años 1931 a 1991. En esta investigación se recomienda que los técnicos de campo realicen las capturas de datos de la forma más precisa, para que esta información pueda ser evaluada más exacta y que estos estudios puedan hacerse en tiempo real con modelos climáticos que caractericen y calculen los parámetros de sequías meteorológicas.
\smallskip

En esta investigación tiene como objetivos comparar tres métodos geoestadístico para la interpolación de datos faltantes de precipitación. Realizando tres pruebas estadistica se determina cual método estima un valor real con el mínimo error posible, para luego completar la serie de datos de precipitación. La estimación espacial de sequías meteorológicas y sus respectivos parámetros, se realiza por medio del SPI a escala de análisis mensual y trimestral, para la cuenca hidrográfica del río Coello presentados entre el período 1996 al 2005.

	\section{\normalsize{METODOLOGÍA}}
La \textcolor{blue}{Figura 1} pertenece a la cuenca hidrográfica del río Coello; ilustra el relieve y ubicación de las estaciones meteorológicas del Instituto de Hidrología, Meteorología y Estudios Ambientales (IDEAM) utilizada en esta investigación, en la escala temporal 1996 al 2005.
		
		\subsection{Área de estudio}
La cuenca hidrográfica del río Coello se encuentra ubicada en la región noroeste del departamento del Tolima - Colombia, entre las coordenadas 4\textsuperscript{o}39\textsuperscript{'}24.56\textsuperscript{''} Norte y 75\textsuperscript{o}19\textsuperscript{'}44.59\textsuperscript{''} Oeste parte alta y 4\textsuperscript{o}17\textsuperscript{'}21.93\textsuperscript{''} Norte y 74\textsuperscript{o}53\textsuperscript{'}30.95\textsuperscript{''} Oeste parte baja, en la vertiente oriental de la cordillera central.

\end{multicols}

\begin{figure}[h]\centering
\centering
\includegraphics[width=400pt]{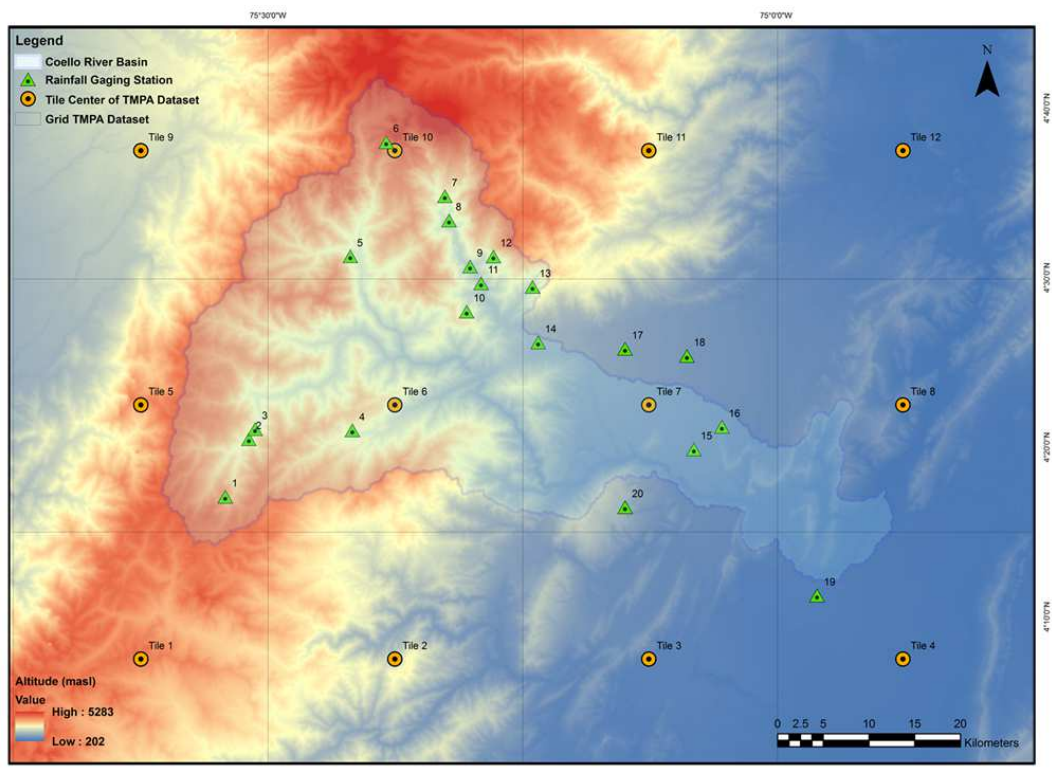}
\captionof{figure}{Distribución espacial de las estaciones meteorológicas del IDEAM de la cuenca hidrográfica del río Coello.}
\end{figure}

\begin{multicols}{2}

Las principales subcuencas del río Coello son los ríos Combeima, Toche, Bermellón, Anaime y Coello Cocora; nace a los 5300 m de altitud y desemboca en el río Magdalena a 280 m de altitud; estas características brindan diversos climas y zonas de vida por la altitud sobre el nivel del mar. El río Coello es la tercer cuenca más grande del departamento y abastece uno de los sistemas de riego más importante del país ($USOCOELLO$). En general la cuenca Coello se evidencian alteraciones por diferentes actividades antropogénicas como agricultura, ganadería, industria, extracción de material de arrastre, entre otros \textcolor{blue}{(Guevara, 2014)}.


		\subsection{Variables y procesamientos de datos}
	\subsubsection{Variable primaria}
Los datos meteorológicos utilizados son de precipitación (Pp). Estos datos fueron suministrados en valores diarios entre el 01 de enero de 1996 al 31 de diciembre de 2005 (10 años), de la red de estaciones meteorológicas activas del IDEAM, distribuidas espacialmente en la cuenca hidrográfica del río Coello \textcolor{blue}{(Tabla 1)}. Las estaciones 1 al 16 se encuentran dentro de la cuenca y las estaciones 17 al 20 se ubican por fuera de la cuenca.

\begin{table}[H] 
\centering \tiny{
{\captionof{table}{Información geográfica de las 20 estaciones meteorológicas del IDEAM, en el sistema de coordenadas WGS-84.}}
        \begin{tabular}{clccllc}
\toprule
\textbf{Est.} & \textbf{Nombre y Cod.IDEAM} & \textbf{Altitud} & \textbf{Lat.} & \textbf{Long.} & \textbf{$\overline{Pp}$ (mm)}\\
\midrule
1 & Cascada [21210150] & 3080 & 4.28472 & -75.54194 & 1293,77\\
2 &	Cucuana Hda [21215130] & 2120 & 4.34136 & -75.51855 & 1139,12\\
3 & Plan [21210140] & 2050 & 4.3512 & -75.51247 & 1209,22\\
4 & Palogrande Hda [21210170] & 2200 & 4.35 & -75.41666 & 1637,94\\
5 & Toche [21210180] & 2000 & 4.52105 & -75.41872 & 1117,46\\
6 & Silencio [21210260] & 2500 & 4.63333 & -75.38333 & 1620,26\\
7 & Palmar [21210220] & 2200 & 4.58011 & -75.32566 & 1321,95\\
8 & Juntas [21210020] & 1765 & 4.55630 & -75.32163 & 1445,66\\
9 & Pastales [21210030] & 1505 & 4.51105 & -75.30083 & 1956,27\\
10 & Darien [21210160] & 1920 & 4.46666 & -75.30427 & 2064,40\\
11 & Secreto [21210080] & 1490 & 4.4945 & -75.29005 & 1832,24\\
12 & Placer [21210110] & 2170 & 4.52094 & -75.27788 & 1861,90\\
13 & Esmeralda [21210120] & 1965 & 4.49091 & -75.23941 & 1575,55\\
14 & Interlaken [21210240] & 1210 & 4.43638 & -75.23388 & 2191,21\\
15 & Cementos Dia. [21215140] & 780 & 4.33130 & -75.08066 & 1354,61\\
16 & Aceituno [21220050] & 680 & 4.35344 & -75.05305 & 1354,97\\
17 & Apto Perales [21245040] & 928 & 4.43011 & -75.14841 & 1676,88\\
18 & Perales Ha.Op. [21245010] & 750 & 4.42313 & -75.08747 & 1463,24\\
19 & Nataima [21185020] & 431 & 4.18761 & -74.959 & 1506,89\\
20 & Resaca [21210190] & 1250 & 4.2745 & -75.14841 & 1897,75\\
\bottomrule
\end{tabular}} \end{table}

\textbf{Preprocesamiento de datos de precipitación}
\medskip

Se desarrolló un $script$, programado en el software estadístico de análisis numérico $GNU-QtOctave$, en donde se transformaron datos diarios de precipitación a acumulados mensuales. El $script$ se configuró para cuando falten el 10\% de los datos se clasifica como NaN, porque no son suficientemente representativos para calcular un valor confiable del promedio acumulado mensual.
\medskip

En esta investigación los valores NaN se estimaron numéricamente por el Método de Radio Normal (MRN) para obtener la base de datos completa con algunos valores estimados estadísticamente. Este método \textcolor{blue}{(Ec.1)} considera la media aritmética de los valores de precipitación de un mes en común, sumados y promediados para estimar el valor faltante $P_{test}$, sin tener en cuenta la distribución espacial de las estaciones \textcolor{blue}{(Grijsen $et$ $al$, 1999)}.

\begin{eqnarray}
P_{test}=\frac{1}{M} \cdot (\frac{N_{test}}{N_{base,1}} \cdot P_{base,1}+\frac{N_{test}}{N_{base,2}} \cdot P_{base,2}+ \nonumber\\
\frac{N_{test}}{N_{base,3}} \cdot P_{base,3}+...+\frac{N_{test}}{N_{base,M}} \cdot P_{base,M})
\end{eqnarray}
Donde:

N\textsubscript{test} = precipitación anual promedio en la estación en estudio

N\textsubscript{base,I} = precipitación anual normal en las estaciones adyacentes (para i=1 a M).
\medskip
\medskip

Una vez completos los datos de precipitación de las 20 estaciones meteorológicas, se procedío a estimar el coeficiente de correlación lineal de Pearson, para ver estadísticamente cuales estaciones se asemejan numéricamente \textcolor{blue}{(Fig.2)}. Para la correlación se establecio que los grupos se conformaran cuando las estaciones tiene un coeficiente de Pearson igual o mayor que 0.5. De acuerdo a lo anterior las estaciones 1, 2 y 3 conforman el Grupo 1; estaciones 4 y 5 como el Grupo 2; estaciones del 6 al 13 como el Grupo 3 y las estaciones del 14 al 20 como el Grupo 4. 

\begin{figure}[H]
\centering
\includegraphics[width=180pt]{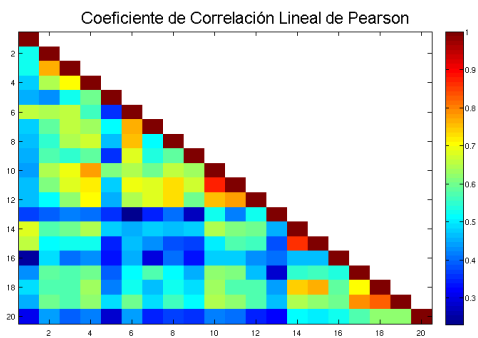}
\captionof{figure}{Correlación lineal entre las precipitaciones promedio multianual de las estaciones de estudio.}
\end{figure}

Con las 20 estaciones integradas en 4 grupos correlacionados numéricamente, se procedío a realizar el análisis de consistencia de datos por medio de las Curvas de Doble Masa (CDM) \textcolor{blue}{(Fig.3)}. Con los valores de precipitación acumulada mensual se calculan los valores acumulados de cada Estación (eje y) versus el promedio acumulado de todas las estaciones del Grupo (eje x); seguidamente se grafica y se identifican las pendientes ($\alpha_{1}$ y $\alpha_{2}$) de los registros meteorológicos. Para esta investigación se determinó que las series de datos de precipitación más recientes (las pendientes correspondientes a $\alpha_{2}$) por ser consideras más confiables y precisos por la mejora tecnológica en los equipos de medición, sirven para ajustar los valores antiguos (correspondientes a $\alpha_{1}$) por el método de CDM.

\begin{figure}[H]
\centering
\includegraphics[width=220pt]{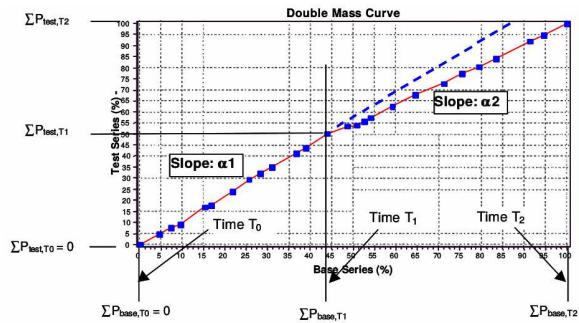}
\captionof{figure}{Esquema que exhibe dos pedientes típicas en los análisis de CDM \textcolor{blue}{(Grijsen $et$ $al$, 1999)}.}
\end{figure}

Para calcular el valor de las pendientes, en el caso de $\alpha_{1}$  se realiza la sumatoria de los valores tanto en $y$ (Test,\textsubscript{T1}) como en $x$ (Base,\textsubscript{T1}), hasta donde acotan las coordenas de la primer pendiente, luego se divide el valor de $y$ sobre $x$ \textcolor{blue}{(Ec.2)}. Para el $\alpha_{2}$ se realiza la sumatoria de los valores de toda la pendiente $y$ (Test,\textsubscript{T2}) menos la pendiente de $y$ (Test,\textsubscript{T1}) dividido entre la sumatoria de la pendiente $x$ (Base,\textsubscript{T2}) menos la pendiente $x$ (Base,\textsubscript{T1}) \textcolor{blue}{(Ec.3)} \textcolor{blue}{(Grijsen $et$ $al$, 1999)}.

\begin{eqnarray}
\alpha_{1}=\frac{\sum_{i=0}^{T\textsubscript{1}}P\textsubscript{Test,i}}{\sum_{i=0}^{T\textsubscript{1}}P\textsubscript{Base,i}} \\
\nonumber\\
\alpha_{2}=\frac{\sum_{i=T\textsubscript{0}}^{T\textsubscript{2}}P\textsubscript{Test,i} - \sum_{i=T\textsubscript{0}}^{T\textsubscript{1}}P\textsubscript{Test,i}}{\sum_{i=T\textsubscript{0}}^{T\textsubscript{2}}P\textsubscript{Base,i} - \sum_{i=T\textsubscript{0}}^{T\textsubscript{1}}P\textsubscript{Base,i}} \\
\nonumber\\
P_{corr j}= P_{test j} \cdot \frac{\alpha_{2}}{\alpha_{1}}
\end{eqnarray}

Obtenidos los valores de $\alpha_{1}$ y $\alpha_{2}$, se calcula el factor de corrección \textcolor{blue}{(Ec.4)}, dividiendo $\alpha_{2}$ entre $\alpha_{1}$, para luego miltiplicarlo por cada uno de los valores de precipitación en donde la pendiente es irregular y posiblemente los valores estan sobre o sub estimados \textcolor{blue}{(Grijsen $et$ $al$, 1999)}. Para citar un caso la \textcolor{blue}{Figura 4} toma como ejemplo la estación 2 del Grupo 1, en donde se percibe una concavidad en la pendiente \textcolor{blue}{(Fig.4a)}. Aplicando el factor de correción se evidencia que disminuye la concavidad y la pendiente tiende a ser lineal \textcolor{blue}{(Fig.4b)}.
\medskip

En el preprocesamiento de datos es importante aclarar que el MRN no se puede comparar con los métodos de estimación geoestadísticos. El MRN no incluye el factor espacial y además exige conocer la variable $Ntest$ (precipitación anual promedio) de la \textcolor{blue}{Ecuación 1} para validar el método, entre los valores estimados con los observados que respectivamente pueden ser NaN, utilizados en las pruebas de calidad de métodos. A diferencia, los métodos geoestadísticos tienen la ventaja de estimar un valor de precipitación en sitios donde no hay mediciones, demostrando la versatilidad de estos métodos como el Kriging Ordinario \textcolor{blue}{(Quiring, 2009)}.

\begin{figure}[H]
 \centering
 \subfigure[]{\includegraphics[width=180pt]{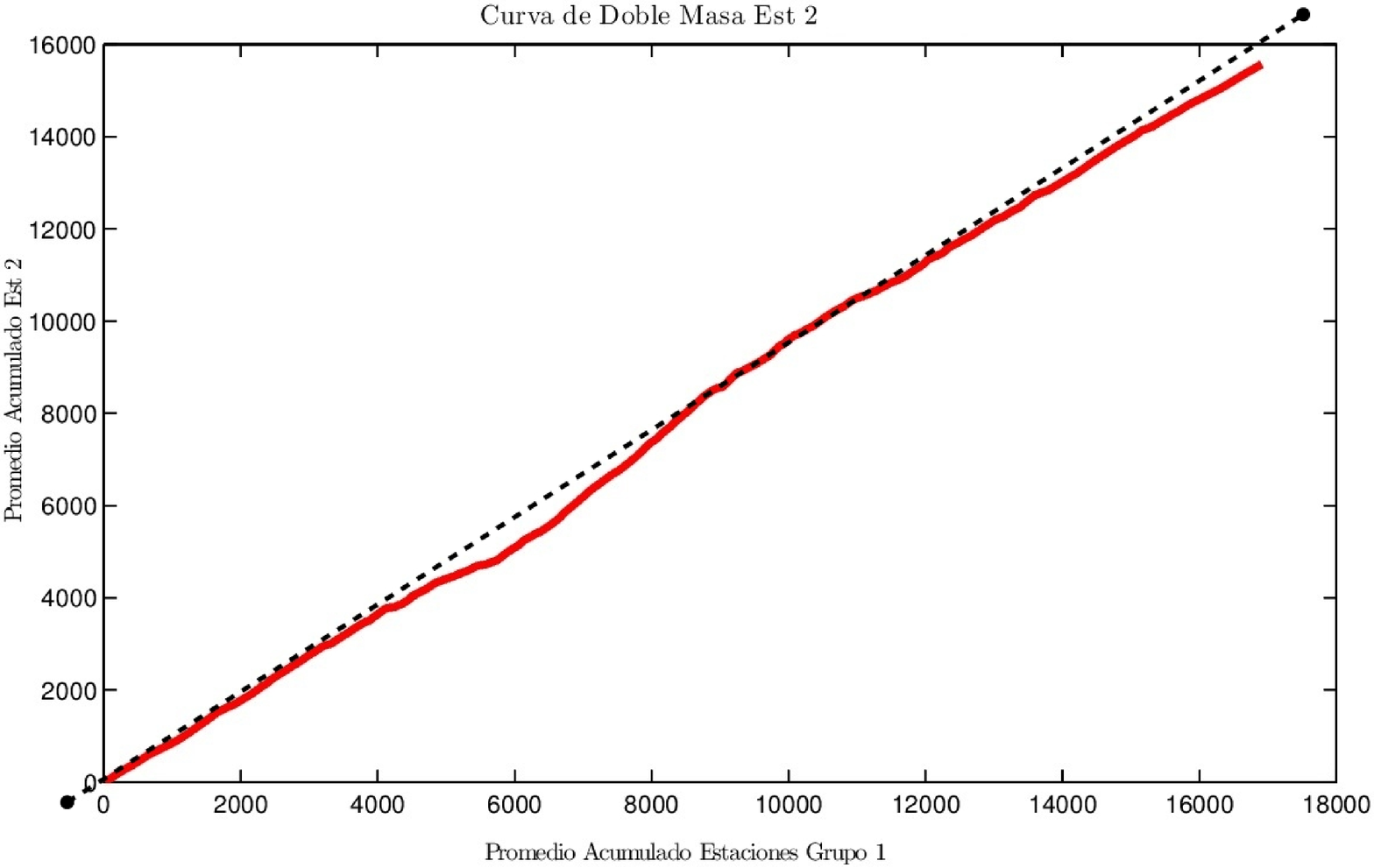}}\\
 \subfigure[]{\includegraphics[width=180pt]{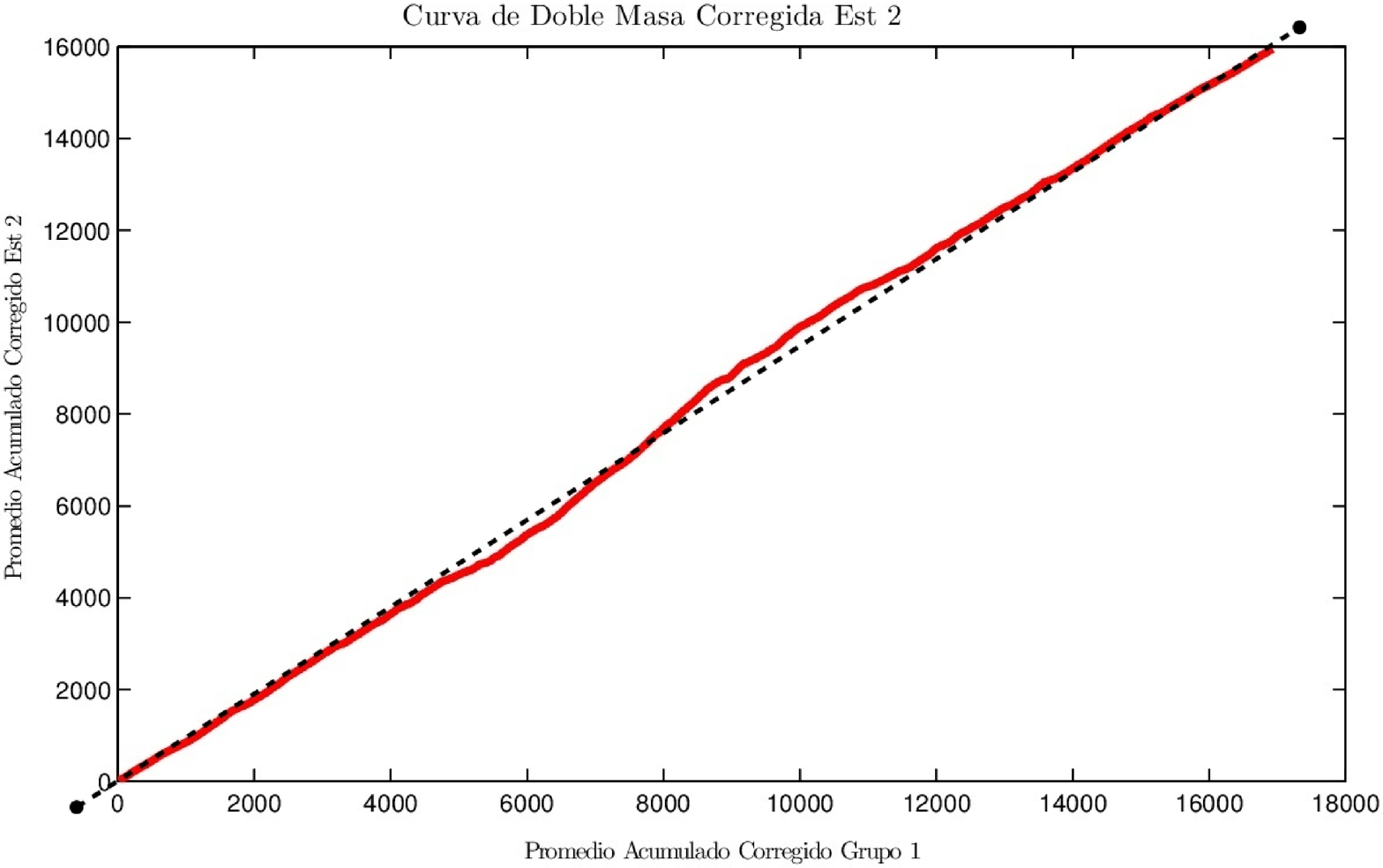}}
 \caption{Análisis de curvas de doble masa de la Estación 2 del Grupo 1, (a) CDM y (b) CDM corregida.}
\end{figure}

	\subsubsection{Variables secundarias}
Las variables secundarias utilizadas en esta investigación fueron la Altitud, la cual es representada en este estudio mediante el DEM SRTM de 3 arcosegundos de la NASA, y datos de lluvia del satélite de la Misión de Medición Lluvias Tropicales (TRMM - Tropical Rainfall Measuring Mission). Estas variables secundarias se utilizaron para comparar métodos de estimación de valores de precipitación por el método de correlación geoestadístico CoKriging Ordinario.
\medskip

La variable DEM corresponde a la Altitud del relieve topográfico de la cuenca del río Coello \textcolor{blue}{(Fig.1)}. Esta variable es importante para comprender las relaciones que tiene la precipitación con la altitud. En la \textcolor{blue}{Figura 5} se grafican las Precipitaciones Anual Multianual de todas las estaciones de estudio (eje $y$) y la altitud de estaciones meteorológicas utilizadas (eje $x$) de la \textcolor{blue}{Tabla 1}, en la cual se evidencia que hay una relación entre el Óptimo Pluviométrico y en este caso \textcolor{blue}{(Fig.5)} ocurre aproximadamente en los 1400 msnm. Es evidente que si se desea comparar métodos geoestadísticos de interpolación de datos faltantes de precipitación, la variable DEM puede ser influyente para estimar valores muy cercanos a los reales en lugares donde no se realizan mediciones, por errores de registro o falta de instrumentación.

\begin{figure}[H]
\centering
\includegraphics[width=210pt]{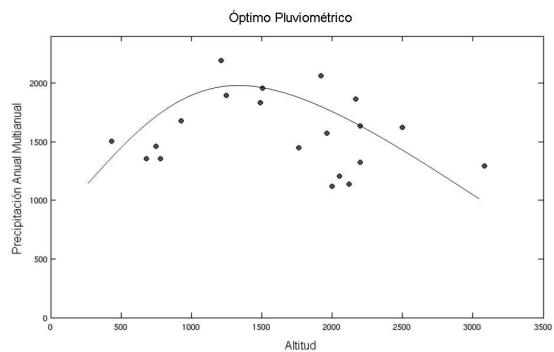}
\captionof{figure}{Comportamiento de la precipitación a la variación de la altitud, Óptimo Pluviométrico de las 20 estaciones de estudio.}
\end{figure}

La variable TRMM corresponde a la precipitación medida satelitalmente, específicamente de la cuenca de estudio y en la escala de tiempo comprendida entre 1998 al 2005. La \textcolor{blue}{Figura 6} representa la cuenca Coello, con las estaciones meteorológicas utilizadas en esta investigación y los puntos de medición, en donde el satélite captura los datos de precipitación. Los datos obtenidos por este satélite son diarios y con el uso de un $Script$ desarrollado en esta investigación y programado en el software estadístico de análisis numérico $MatLab$, se calcularon los datos acumulados mensuales de precipitación satelital para los 20 puntos de medición con 0.25 grados de resolución espacial. Es relevante ejecutar y analizar el método CoKrigin Ordinario utilizando como variable secundaria datos de lluvia medida por el satélite TRMM, para evaluar los mecanismo de medición de datos de precipitación (satelital y pluviométrica) en la estimación de datos faltantes de precipitación.
 
\begin{figure}[H]
\centering
\includegraphics[width=210pt]{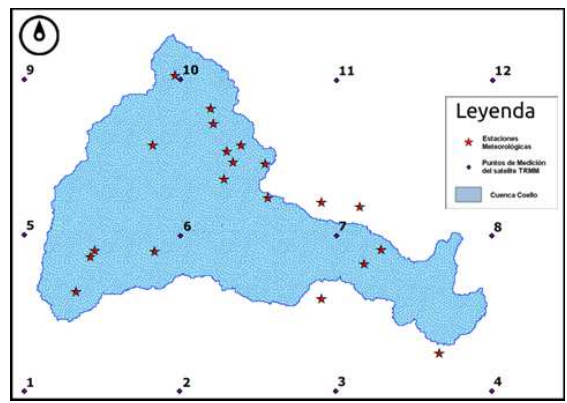}
\captionof{figure}{Puntos de medición de precipitación satelital TRMM.}
\end{figure}

		
		\subsection{Métodos de interpolación geoestadística}
Los métodos de interpolación geoestadísticos sirven para estimar númericamente un valor puntual no conocido utilizando algunos valores vecinos conocidos, asociado con una variable geográfica \textcolor{blue}{(Giménez-Palomares \& Cuador-Gil, 2014)}. \textcolor{blue}{Bayat $et$ $al$ (2014)} define el Kriging Ordinario como el mejor estimador lineal insesgado de la variable no muestreada.

	\subsubsection{Kriging y CoKriging Ordinario}
Para estimar valores con el método de Kriging es necesario calcular los Semivariogramas Experimentales Direccionales, en donde los valores de precipitación se ajustan a una distribución estadística. En este proceso se eliminan los valores estimados por el método estadístico de MRN de la \textcolor{blue}{ecuación 1} reemplazados por el símbolo inicial NaN.
\medskip

En el caso de estudio las distribuciones Gaussianas y Exponenciales son las más comunes en los valores de precipitación de esta investigación. La configuración para calcular los semivariogramas experimentales dirrecionales generalmente se utilizaron ángulos entre 0\textsuperscript{o} y 160\textsuperscript{o}, una tolerancia de 45 y número de Lag's  de 12 a distancias de 3 unidades.
\medskip

Para esta investigación se utilizó el software $ArcGIS$, usando la herramienta Ánalisis Geoestadístico. Como ejemplo la \textcolor{blue}{Figura 7} muestra el semivariograma experimental dirrecionado a 135\textsuperscript{o} de los valores promedios multianuales del mes de noviembre ajustados a una distribución Gaussiana, donde el eje $y$ representa la Variación y el eje $x$ los Números de Lag's.

\begin{figure}[H]
\centering
\includegraphics[width=240pt]{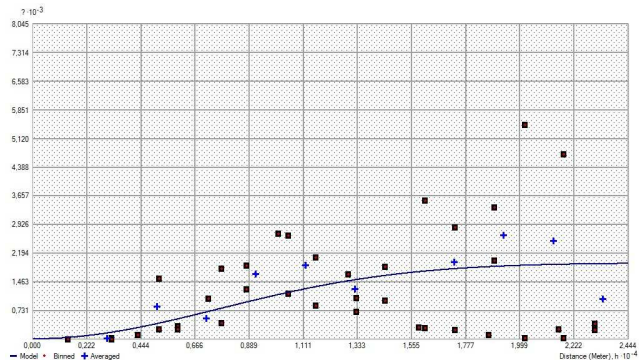}
\captionof{figure}{Semivariograma del promedio multianual de noviembre de todas las estaciones de estudio.}
\end{figure}

Una ves calculado el semivariograma se calcula el estimador Kriging Ordinario, que tiene como forma la \textcolor{blue}{Ecuación 5}:

\begin{eqnarray}
z^{*}(\textbf{u}) = \sum_{\alpha=1}^{n}  \lambda_{\alpha}  z(\textbf{u}_{\alpha})
\end{eqnarray}
Donde:

$z^{*}$ corresponde al atributo desconocido a estimar en un punto de coordenadas \textbf{u}, que a partir de $n$ valores conocidos de $z$, cuyas coordenas son $\textbf{u}_{\alpha}$, con $\alpha = 1$, ... n. El valor de $\lambda_{\alpha}$ (\textbf{u}) son coeficientes de ponderación desconocidos \textcolor{blue}{(Cassiraga, 2008)}.
\medskip

Para esta investigación se compararon tres métodos de interpolación geoestadísticos: (i) Kriging Ordinario [OK], (ii) CoKriging Ordinario + Modelo de elevación Digital [CoK+DEM] y (iii) CoKriging + lluvia satelital TRMM [CoK+TRMM]. Los resultados de estos métodos en el software $ArcGIS$ son mapas formato ASCII de toda la cuenca, en un mes específico, en donde se encuentran los valores estimados según el Kriging o CoKriging.
\medskip

En cada uno de los métodos se realiza el proceso de verificación que consiste en eliminar 3 estaciones representativas y con datos conocidos (en esta investigación fueron las Est 2, 8 y 16 descritas en la \textcolor{blue}{Tabla 1}) para comparar los valores estimados con los reales de todas las estaciones; de esta forma validar los resultados de estos métodos para estimar datos faltantes de precipitación \textcolor{blue}{(Tabla 2)}. La escala temporal de los datos de lluvia satelital TRMM corresponden a los disponibles cuando el satélite comenzó sus mediciones (01 de enero de 1998) \textcolor{blue}{(Serrat-Capdevila $et$ $al$, 2014)}.

\begin{table}[H] \centering \tiny{
{\captionof{table}{Descripción de los métodos de interpolación geoestadísticos comparados en esta investigación.}}
        \begin{tabular}{llcc}
\toprule
\textbf{Método} & \textbf{Datos} & \textbf{Escala temporal} &  \textbf{N. Est}\\
\midrule
OK & Pp & 10 años & 20\\
OK verificación & Pp & 10 años & 17\\
CoK+DEM & Pp+Altitud & 10 años & 20\\
CoK+DEM verificación & Pp+Altitud & 10 años & 17\\
CoK+TRMM & Pp+Pp(Satelital) & 8 años & 20\\
CoK+TRMM verificación & Pp+Pp(Satelital) & 8 años & 17\\
\bottomrule
\end{tabular}} \end{table}

		\subsection{Pruebas de calidad de métodos geoestadísticos}
Para validar y seleccionar el mejor método de estimación de datos se realizan tres pruebas de calidad de métodos geoestadisticos: (i) Raíz del Error Cuadrático Medio (Root Mean Squared Error - RMSE) usando el método de Validación Cruzada en las estaciones de verificación 2, 8 y 16; (ii) Criterio de Información de Akaike (Akaike Information Criterion - AIC) y (iii) Criterio de Información Bayesiano (Bayesian Information Criterion - BIC). En nuestro caso se comparan los valores estimados versus los observados de precipitación en los tres métodos OK, CoK+DEM y CoK+TRMM; estableciendo por medio de estas pruebas, cual método geoestadístico estima valores con el mínimo error.

	\subsubsection{Prueba RMSE}
Basado en la \textcolor{blue}{Ecuación 6}, RMSE es una medida de desempeño, que cuantifica la diferencia entre los valores estimados y los valores observados, el método con mejor ajuste corresponde a un menor valor de RMSE \textcolor{blue}{(Bayat $et$ $al$, 2014)}:

\begin{eqnarray}
RMSE = \sqrt{\frac{1}{n} \cdot \sum_{i=1}^{n} (X_{o} - {X}_{e})^{2}}
\end{eqnarray}
Donde:

$n$ = Número de predicciones

$X_{o}$ = Valores observados

$X_{e}$ = Valores estimados.

	\subsubsection{Prueba AIC}
Basado en la \textcolor{blue}{Ecuación 7}, AIC es una medida de calidad relativa de un método geoestadístico para un conjunto de datos, en donde el método con mejor ajuste corresponde a un menor valor de AIC \textcolor{blue}{(Pan, 2001)}:

\begin{eqnarray}
AIC = n \cdot ln(RMSE) + 2 \cdot K
\end{eqnarray}
Donde:

$n$ = Número de valores observados

$ln$ = Logaritmo Natural

$K$ = Número de parámetros en el método.

$RMSE$ = Raíz del error cuadrático medio \textcolor{blue}{Ec.6}.

	\subsubsection{Prueba BIC}
Basado en la \textcolor{blue}{Ecuación 8}, BIC es una función de bondad para comparar las métodologías, en donde el método con mejor ajuste corresponde a un menor valor de BIC \textcolor{blue}{(Posada \& Noguera, 2007)}.

\begin{eqnarray}
BIC = n \cdot ln(RMSE) + K \cdot ln(n)
\end{eqnarray}
Donde:

$n$ = Número de valores observados

$ln$ = Logaritmo Natural

$K$ = Número de parámetros en el método

$RMSE$ = Raíz del error cuadrático medio \textcolor{blue}{Ec.6}.

	\subsection{Algoritmo e Interpretación del SPI}
El Índice de Precipitación Estandarizado o SPI (Standardised Precipitation Index) desarrollado por \textcolor{blue}{Mckee $et$ $al$ (1993)} cuantifica y clasifica el déficit de precipitación para múltiples escalas de tiempo (1, 3, 6, 12, 24, 48 meses) \textcolor{blue}{(Ec.9)}, basado en la probabilidad acumulada de alguna serie de datos de precipitación que se midieron en un lugar determinado \textcolor{blue}{(Tabla 3)} \textcolor{blue}{(Gocic \& Trajkovic, 2014)}.

\begin{eqnarray}
x_{i} = \frac{X_{i} - \bar{X}}{S{x}}
\end{eqnarray}
Donde:

$x\textsubscript{i}$ = Serie de precipitación estandarizada

X\textsubscript{i} = Serie de datos de precipitación

$\bar{X}$ = Media aritmética de los datos

$S\textsubscript{x}$ = Desviación estandar de la serie de datos.

\begin{table}[H] \centering \tiny{
{\captionof{table}{Descripción de las categorías del SPI, tomado y adaptado de: \textcolor{blue}{Mckee $et$ $al$ (1995)}.}}
        \begin{tabular}{clcc}
\toprule
\textbf{Categoría} & \textbf{Descripción} & \textbf{Valor SPI} &  \textbf{Probabilidad}\\
\midrule
D1 & Normal & -0.99 $\leq$ SPI $<$ 0.0 & 60\%\\
D2 & Moderadamente Seco & -1.0 $\leq$ SPI $<$ -1.49 & 40\%\\
D3 & Seco & -1.5 $\leq$ SPI $<$ -1.99 & 20\%\\
D4 & Muy Seco & SPI $<$ -2.0 & 10\%\\
\bottomrule
\end{tabular}} \end{table}

Para esta investigación se desarrollo un $Script$ para el calculo de SPI, ejecutado en el software $MatLab$, específicamente programado para las escalas mensuales y trimestrales.

	\subsubsection{Parámetros de sequías de SPI}

La gráfica de los valores del SPI en una escala de tiempo permiten calcular los parámetros de sequías, como la severidad, duración, intensidad y frecuencia \textcolor{blue}{(Fig.8)}. La duración (inicio y termino del evento de sequía) y frecuencias (números de eventos) se calculan por medio del análisis en la gráfica, revelando los patrones temporales de las sequías meteorológicas \textcolor{blue}{(Mishra \& Singh, 2010)}. La severidad de las sequías meteorológicas se define como la sumatorias de los valores de las áreas bajo la curva, en donde los valores de SPI se encuentran por debajo de la media aritmética y la intensidad es el valor más negativo que se presenta en una sequía ilustrada en la gráfica del SPI \textcolor{blue}{(Sirdaş \& Sen, 2003)}:

\begin{figure}[H]
\centering
\includegraphics[width=230pt]{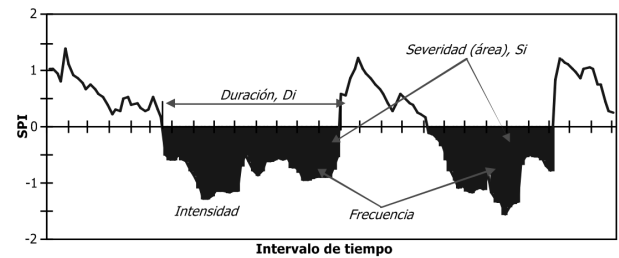}
\captionof{figure}{Características y parámetros de sequías de SPI, tomado y adaptado de: \textcolor{blue}{Valdés $et$ $al$ (2008)}.}
\end{figure}


	\section{\normalsize{RESULTADOS}}
	\subsection{Método Geoestadístico de Estimación}
	
Los parámetros numéricos de entrada para las tres pruebas de calidad de métodos geoestadísticos se exponen en la \textcolor{blue}{Tabla 4}. Las pruebas se ejecutaron en los tres métodos de estimación de datos faltantes de precipitación evaluados, considerados como Verificación, utilizando las estaciones meteorológicas 2, 8 y 16. La variable $K$ corresponde al número de parámetros del método ejecutados en el software $ArcGIS$, utilizados en las pruebas AIC y BIC. Para el caso fueron cinco: Angle, Bandwidth, Type Distribution, Lag size y Number of Lag's.

\begin{table}[H] \centering \tiny{
{\captionof{table}{Valores de parámetros de entrada en las pruebas de calidad de modelos geoestadísticos.}}
        \begin{tabular}{clcccccc}
\toprule
\textbf{Prueba} & \textbf{Método} & \textbf{Est.} & \textbf{$n$} & \textbf{$K$} & \textbf{$RMSE$} & \textbf{$Valor$} & \textbf{$Media$}\\
\midrule

\multirow{9}{0.3cm}{RMSE}
&  & 2 & 36 &  & 82, 01 &  \\ & OK & 8 & 96 &  & 94,76 &  & 99,65 \\ &  & 16 & 96 &  & 122,18 &  \\ \cmidrule{2-8}
&  & 2 & 36 &  & 43,91 & \\ & * CoK+DEM & 8 & 96 &  & 93,14 & & 86,60 * \\ &  & 16 & 96 &  & 122,74 & \\ \cmidrule{2-8}
&  & 2 & 36 &  & 94,35 & \\ & CoK+TRMM & 8 & 96 &  & 96,03 & & 107,25 \\ &  & 16 & 96 &  & 131,36 & \\ \cmidrule{1-8}

\multirow{9}{0.3cm}{AIC}
&  & 2 & 36 & 3 & 82,01 & 164,64 \\ & OK & 8 & 96 & 3 & 94,76 & 442,93 & 358,30 \\ &  & 16 & 96 & 3 & 122,18 & 467,32 \\ \cmidrule{2-8}
&  & 2 & 36 & 5 & 43,91 & 146,16 \\ & * CoK+DEM & 8 & 96 & 5 & 93,14 & 445,28 & 354,40 * \\ &  & 16 & 96 & 5 & 122,74 & 471,76 \\ \cmidrule{2-8}
&  & 2 & 36 & 5 & 94,35 & 173,69 \\ & CoK+TRMM & 8 & 96 & 5 & 96,03 & 448,21 & 366,73 \\ &  & 16 & 96 & 5 & 131,36 & 478,28 \\ \cmidrule{1-8}

\multirow{9}{0.3cm}{BIC}
&  & 2 & 36 & 3 & 82,01 & 169,40 \\ & * OK & 8 & 96 & 3 & 94,76 & 450,63 & 365,01 * \\ &  & 16 & 96 & 3 & 122,18 & 475,02 \\ \cmidrule{2-8}
&  & 2 & 36 & 5 & 43,91 & 154,08 \\ & CoK+DEM & 8 & 96 & 5 & 93,14 & 458,10 & 365,59 \\ &  & 16 & 96 & 5 & 122,74 & 484,58 \\ \cmidrule{2-8}
&  & 2 & 36 & 5 & 94,35 & 181,61 \\ & CoK+TRMM & 8 & 96 & 5 & 96,03 & 461,03 & 377,91 \\ &  & 16 & 96 & 5 & 131,36 & 491,10 \\

\bottomrule
\end{tabular}} \end{table}

El resultado de las pruebas de la \textcolor{blue}{Tabla 4} establece que el método de estimación geoestadístico CoKriging + Modelo de Elevación Digital (CoK+DEM) es el método que estima valores faltantes más aproximados a las condiciones de la cuenca, según los valores de las pruebas RMSE (86,60) y AIC (354,40). La prueba BIC favorece al método OK (365,01) pero se encuentra muy cercano al CoK+DEM (365,59), de lo cual se infiere que en general las tres pruebas favorecen esta metodología geoestadística para estimar datos de precipitación.

	\subsection{Análisis del SPI}
	
Aplicando el método seleccionado se estimaron los valores de lluvia y se calcularon los valores de SPI a escalas de análisis mensual y trimestral para los 10 años de estudio. En la \textcolor{blue}{Figura 9} se representan los valores de SPI para toda la cuenca por medio de un gráfico de cajas. En la \textcolor{blue}{Figura 9a} los meses de Enero y Septiembre contienen valores atípicos a los demás meses del año y en \textcolor{blue}{Figura 9b} corresponden a los meses de Enero, Febrero, Mayo, Septiembre y Diciembre. Estos valores atípicos se alejan de los demás datos de SPI y se asocian a eventos extraordinarios de poca lluvia (negativos) y periodos muy húmedos (positivos). 

\end{multicols}

\begin{figure}[H] \centering
 \begin{multicols}{2}
\includegraphics[width=242pt]{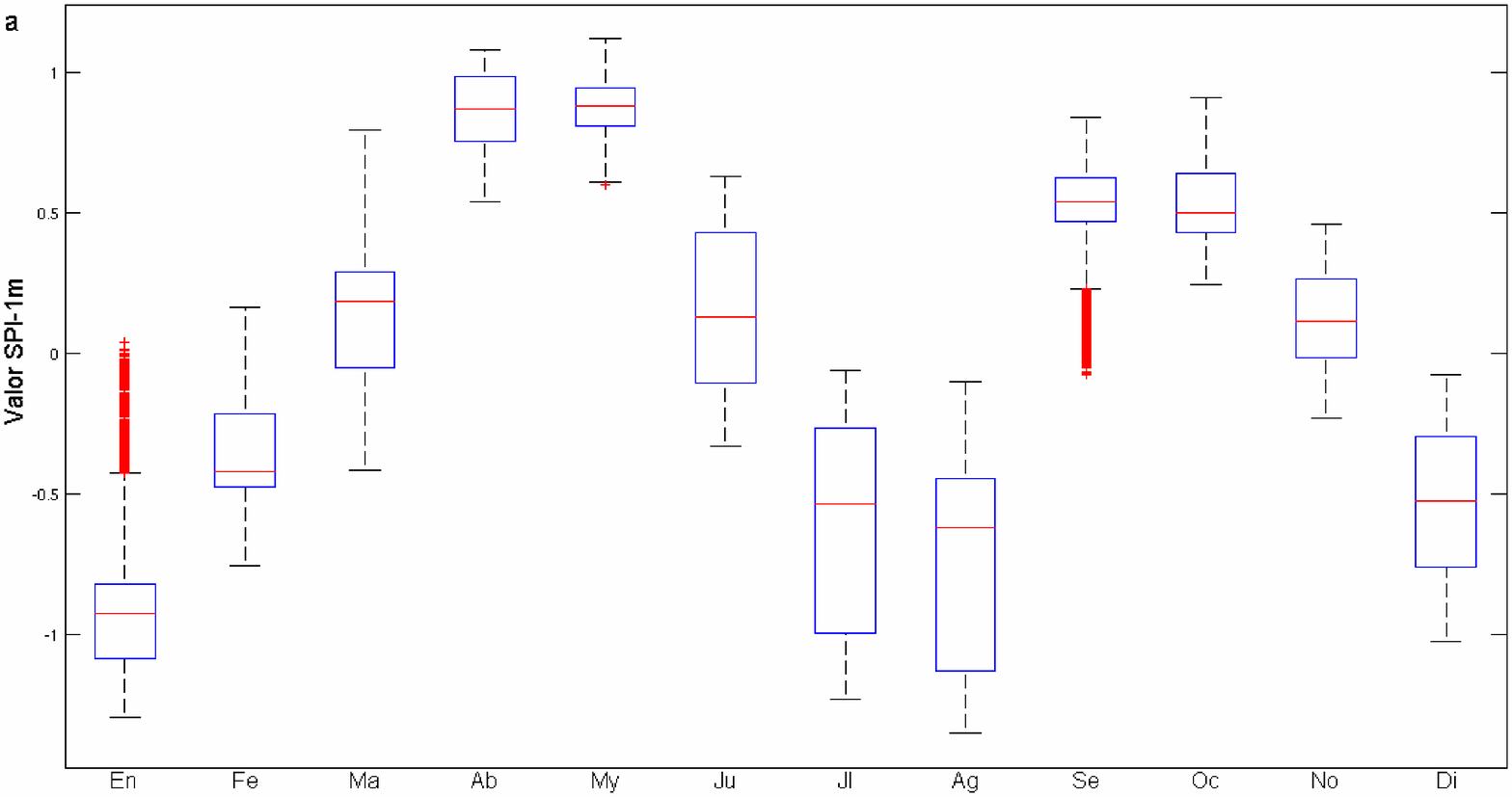}
\includegraphics[width=250pt]{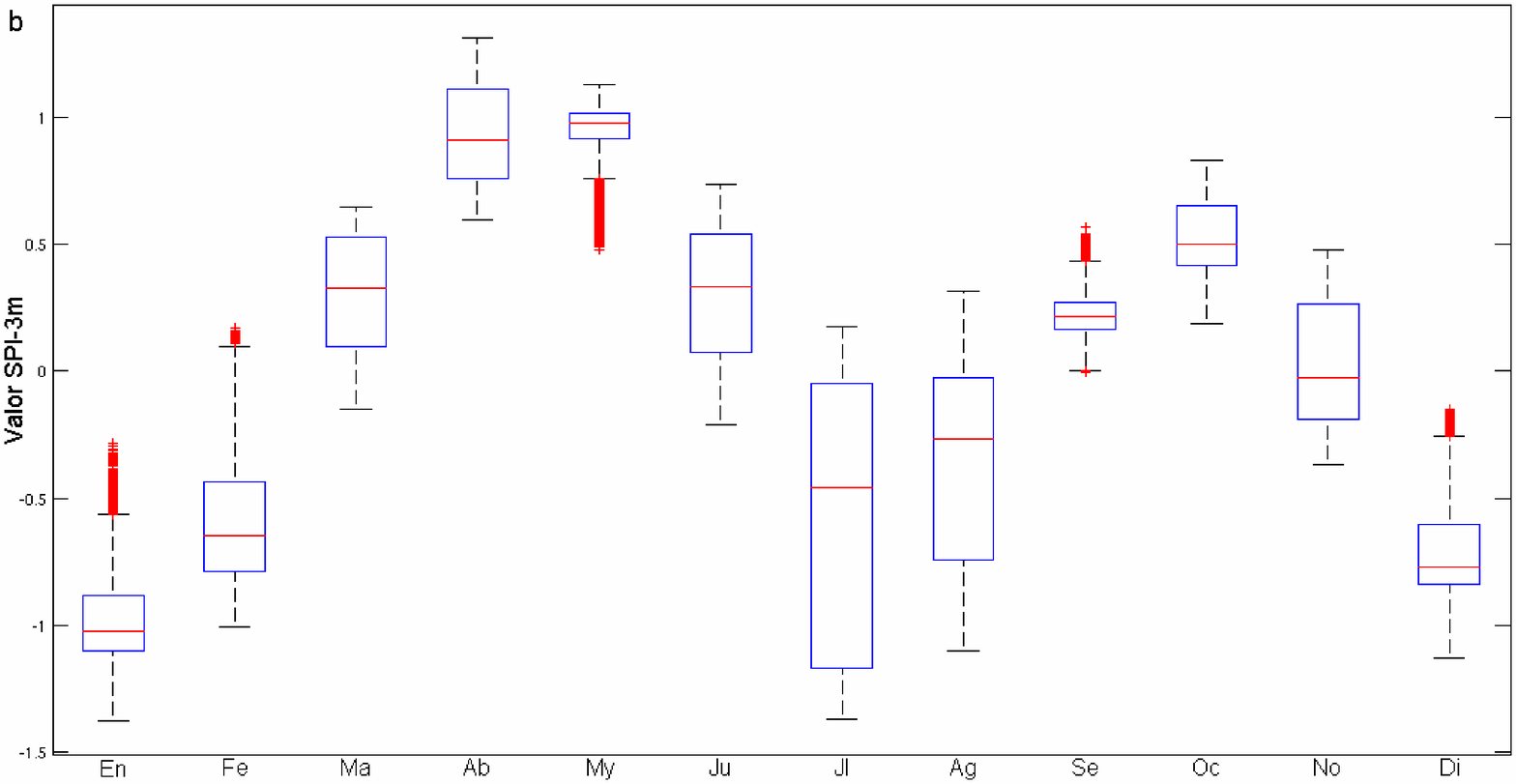}
\end{multicols}
\caption{Medidas de tendencia central, dispersión y simetría de los valores de SPI de cada mes, de la cuenca del río Coello; (a) SPI escala de análisis mensual y (b) SPI escala de análisis trimestral.}
\end{figure}

\begin{multicols}{2}

En general los valores de SPI graficados en ambas escalas de análisis demuestran que son asimétricos y heterogéneos por que la Mediana tiende a concentrarse en a los extremos de las cajas. El mes de Julio y Agosto en ambas figuras demuestran una amplia dispersión en los datos de SPI, de igual forma estos meses poseen los valores más negativos de todo el año. Este comportamiento de los valores de SPI en ambas escalas de análisis, demuestra que Julio y Agosto son los meses de mayor íntensidad como se gráfica en los \textcolor{blue}{Apéndice 1} y \textcolor{blue}{2} en donde los colores más azules oscuros de la barra espectral lateral derecha corresponden a valores de SPI más negativos. Espacialmente en estos dos meses, los valores de SPI se concentran en la parte media y baja de la cuenca. En contraste los meses de Enero y Febrero se caracterizan por pertenecer al grupo de los meses más secos del año, con la característica de que los valores negativos se concentran en la parte alta de la cuenca.

\subsection{Caracterización de la severidad de sequías}

En la \textcolor{blue}{Figura 10} se grafican los valores de SPI para toda la cuenca en la serie de tiempo de 10 años de estudio a dos escalas de análisis (mensual y trimestral). Para la gráfica se utilizaron las estaciones meteorológicas ubicadas dentro de la cuenca (de la 1 a la 16), representadas en la \textcolor{blue}{Figura 1}. En la barra lateral derecha de las \textcolor{blue}{Figuras 10 a} y \textcolor{blue}{b} se exponen los intervalos de las categorías de los valores de SPI de la \textcolor{blue}{Tabla 3}. Las regiones que se encuentran por debajo del valor de SPI = 0.0 corresponden a eventos de sequías meteorológicas presentes en la cuenca hidrográfica del río Coello.

\end{multicols}

\begin{figure}[H]
 \centering
 \subfigure[]{\includegraphics[width=450pt,height=140pt]{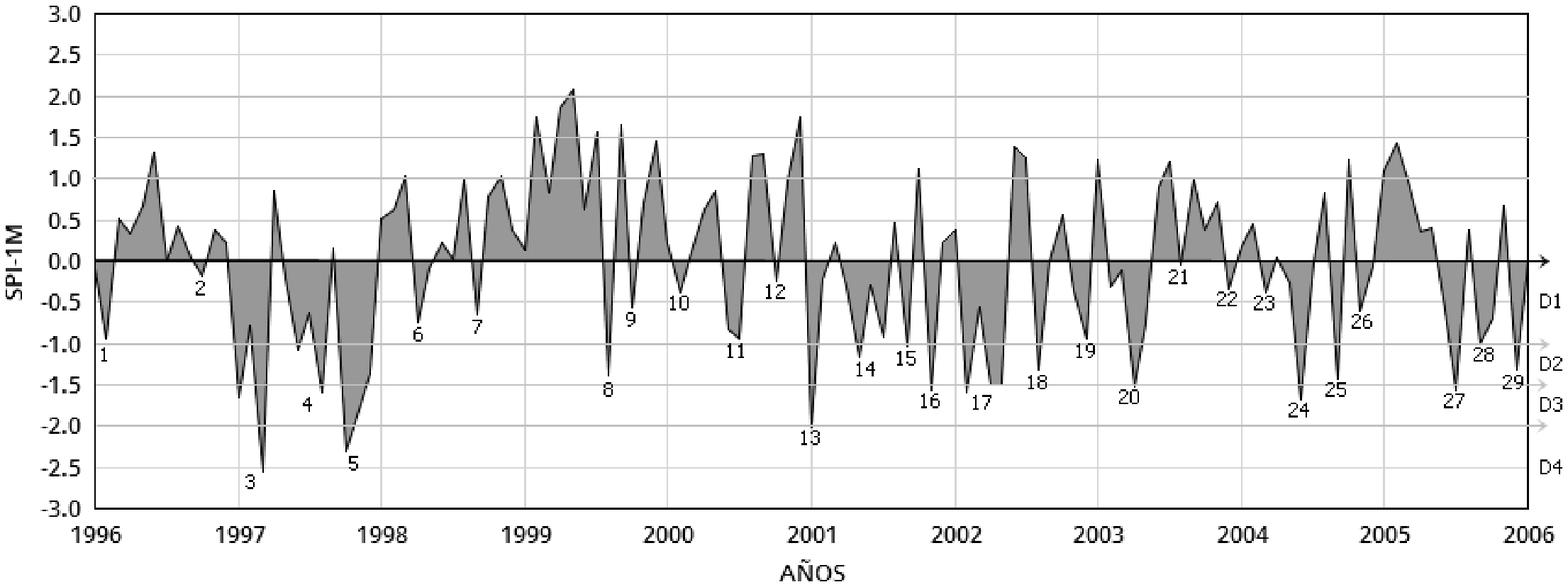}}\\
 \subfigure[]{\includegraphics[width=450pt,height=140pt]{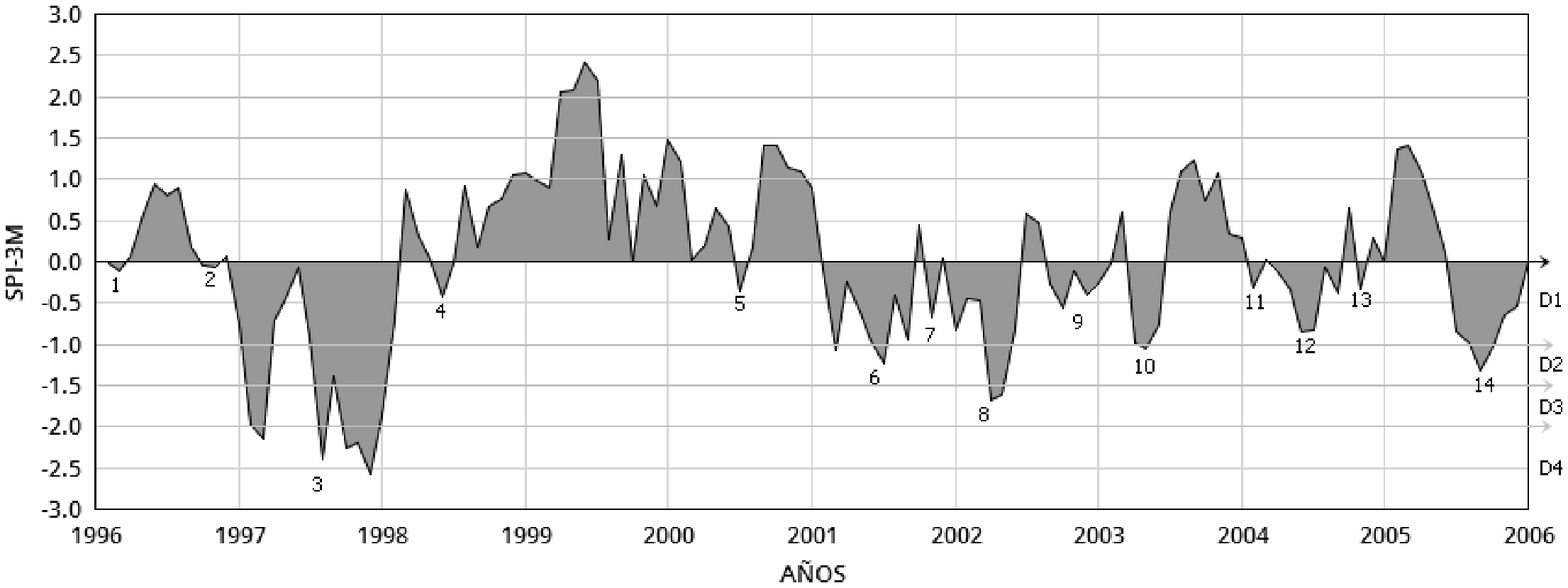}}
 \caption{Dinámica y clasificación de los valores de SPI en la serie de tiempo Enero 1996 a Diciembre 2005 de las estaciones meteorológicas ubicadas dentro de la cuenca [1 al 16]; (a) SPI escala de análisis mensual y (b) SPI escala de análisis trimestral.}
\end{figure}

\begin{multicols}{2}

La caracterización del SPI a escala de análisis mensual de la \textcolor{blue}{Figuras 10 a} se evidencian 29 eventos de sequías. En el \textcolor{blue}{Apéndice 3} se caracterizan las sequías meteorológicas en 10 años de la cuenca hidrográfica del río Coello. En ese intervalo de tiempo se evidencian según las categorías de sequías (ver \textcolor{blue}{Tabla 3}) 13 eventos D1 (Normal), 7 D2 (Moderadamente seco), 6 D3 (Secos) y 3 D4 (Muy secos). Las tres sequías de mayores impactos D4 (N. de Sequía 3, 5 y 13) en la cuenca duraron entre 2 a 4 meses, con una intensidad máxima de -2,55 y una severidad máxima de 5,53; en estas sequías es probable en un 10\% que ocurran 3 eventos en 10 años.
\medskip

La caracterización del SPI a escala de análisis trimestral de la \textcolor{blue}{Figuras 10 b} se evidencian 14 eventos de sequías descritas en el \textcolor{blue}{Apéndice 3}. En el intervalo de tiempo de 10 años se evidenciaron 9 eventos de categoría D1 (Normal), 3 D2 (Moderadamente seco), 1 D3 (Seco) y 1 D4 (Muy seco). La sequía de mayor impacto D4 (N. de Sequía 3) en la cuenca, duró 14 meses, con una intensidad de -2,57 y una severidad máxima de 20,43; esta sequía es probable en un 10\% que ocurra 1 evento en 10 años.


\section{\normalsize{DISCUSIÓN}}

Los métodos geoestadísticos de estimación comparados en esta investigación se basaron en el método Kriging Ordinario, que se caracteriza por proporcionar mejores predicciones utilizando bases de datos relativamente pequeñas que en nuestro caso consiste a 10 años de registros de precipitación mensual, coincidiendo con los plateado por los autores \textcolor{blue}{Schwab \& Marx (2015)}. El OK puede ser una alternativa sencilla a los modelos espaciales más complejos para los pequeños tamaños de muestra. 
\medskip

De tres métodos geoestadísticos comparados en esta investigación, se estable por medio de las pruebas de calidad que el CoK+DEM es el mejor estimador de datos faltantes de precipitación. Los autores \textcolor{blue}{Williams $et$ $al$ (2011)} y \textcolor{blue}{Patel $et$ $al$ (2016)} concuerda con la metodología y resultados de esta investigación, que básicamente recomienda utilizar el método de Cokrigin Ordinario, asociada con datos de un Modelo de Elevación Digital como una variable de predicción para el proceso de Kriging, contribuyendo a mejorar la estimación de valores de precipitación en una serie de tiempo deseada. De igual manera los resultados de los autores \textcolor{blue}{Othman $et$ $al$ (2015)} y \textcolor{blue}{Sirdaş \& Sen (2003)} ilustran las estimaciones mediante el uso de la técnica de interpolación Kriging en mapas temáticos que zonifican las sequías por categoría y facilitan la caracterización de los parámetros. \textcolor{blue}{Bayat $et$ $al$ (2014)} en contraste comparó el método clásico OK con el moderno Bayesian Maximum Entropy o BEM para estimar datos de precipitación en el lago Namak, parte central de Iran, evaluando espacial y temporalmente (1966 al 2005) los patrones de sequías meteorológicas. En él establece superioridad de BME sobre OK, posiblemente puede ser debido a la estructura de las ecuaciones de BME que tiene en cuenta tanto los datos fuertes y blandos de manera simultánea. De los resultados de \textcolor{blue}{Bayat $et$ $al$ (2014)} los autores de esta investigación sugerimos una réplica y comparación a futuro entre el método clásico y moderno para la aproximación a la estimación espacial de sequías en la cuenca del río Coello.
\medskip

Los resultados de esta investigación en conjunto con \textcolor{blue}{Patel $et$ $al$ (2007)} favorecen a la escala de análisis trimestral de SPI; se ha caracterizado por ser más eficaz en la captura de los patrones estacionales de sequías meteorológicas, en nuestro caso representada gráficamente en la \textcolor{blue}{Figura 10 b} y en tabulación en el \textcolor{blue}{Apéndice 3} sección SPI-3M. Esta escala de análisis de SPI revela beneficios de gran utilidad que a su vez contribuyen a mejorar la planificación y gestión del recurso hídrico, como los sugeridos por \textcolor{blue}{Gocic \& Trajkovic (2014)} por medio del cálculo y caracerización de los parámetros de sequías apartir del SPI trimestral. \textcolor{blue}{Vicente-Serrano (2006)} sugiere investigar por medio de análisis comparativos entre los índices de sequía basados en la precipitación y la evapotranspiración para evaluar los impacto en la producción agrícola, forestal y pecuaria para regiones de interés social, ambiental y económico de un país.
\medskip

En la \textcolor{blue}{Figura 11} se ilustra el Índice de El Niño Oceánico (ONI) en el intervalo de tiempo comprendido entre Enero 1996 a Diciembre del 2005 \textcolor{blue}{(Jan, 2016)}. Los resultados de esta investigación concuerdan con la temporalidad de este índice macroclimático. En la \textcolor{blue}{Figura 10 a} y \textcolor{blue}{b} entre los años 1997 y 1998 contiene los valores de SPI más negativos y de mayores afectaciones a la cuenca de estudio, calculados apartir del método geoestadistico CoK+DEM; en estos mismos años el autor \textcolor{blue}{Jan (2016)} gráfica la dinámica y clasificación de los episodios El Niño, acoplándose exactamente a los eventos de sequías meteorológicas de mayores magnitudes para esta investigación.

\end{multicols}

\begin{figure}[H]
\centering
\includegraphics[width=420pt,height=120pt]{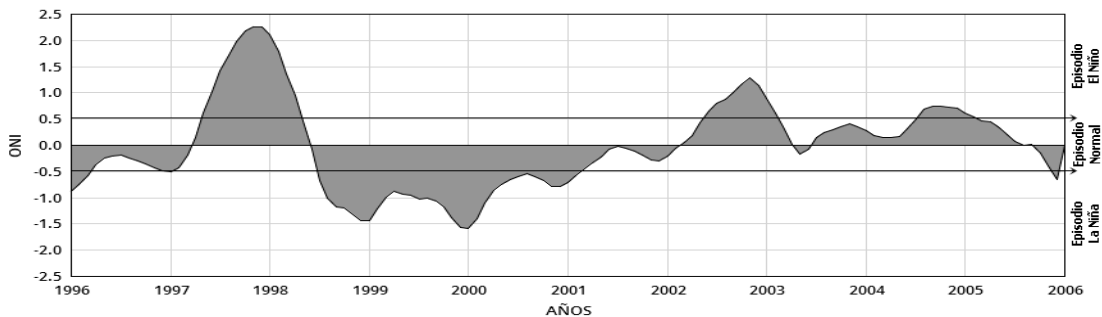}
\captionof{figure}{Índice de El Niño Oceánico (ONI) entre 1996-2005, Elaborado a partir de NOAA-ESRL \textcolor{blue}{(Jan, 2016)}.}
\end{figure}

\begin{multicols}{2}

	\section{\normalsize{CONCLUSIONES}}

El mejor método geoestadístico para estimar datos de precipitación según esta investigación es el CoKriging Ordinario correlacionado con la variable secundaria de la Altitud de un Modelo de Elevación Digital. Los meses más secos del año son Julio y Agosto caracterizados por sequías concentradas en la parte media y baja de la cuenca hidrográfica del río Coello. Los meses de Enero y Febrero de igual manera son meses recurrentes a eventos de sequías en la parte alta de la cuenca. La investigación favorece realizar estudios de SPI a escalas de análisis trimestrales, ya que optimiza la información disponible y facilita el análisis y caracterización de las sequías. Las sequías de mayores afectaciones de la investigación se ajustan directamente a la temporalidad de la dinámica del índice macroclimático ONI (episodio El Niño 1997-1998).

\end{multicols}

\newpage

\bigskip
\textbf{Agradecimientos}
\medskip

Los autores agradecen al Instituto de Hidrología, Meteorología y Estudios Ambientales (IDEAM, Colombia) por suministrar los datos de precipitación; al Consortium for Spatial Information (CGIAR-CSI, Estados Unidos de America) por aprovisionar el Modelo de Elevación Digital de la cuenca de estudio y a la National Aeronautics and Space Administration (NASA - Estados Unidos de America) y Japanese Space Agency (JAXA - Japón) en el proyecto Tropical Rainfall Measuring Mission (TRMM) por permitirnos acceder a los datos de lluvia medidas por este satélite.
\bigskip

\begin{multicols}{2}

\end{multicols}

\newpage
\begin{center}
\textbf{APÉNDICE 1}
\end{center}
\bigskip
\bigskip
\bigskip
\bigskip
{Mapas de SPI a escala mensual para la cuenca del río Coello.}

\begin{figure}[H]
 \begin{multicols}{3}
\includegraphics[width=160pt]{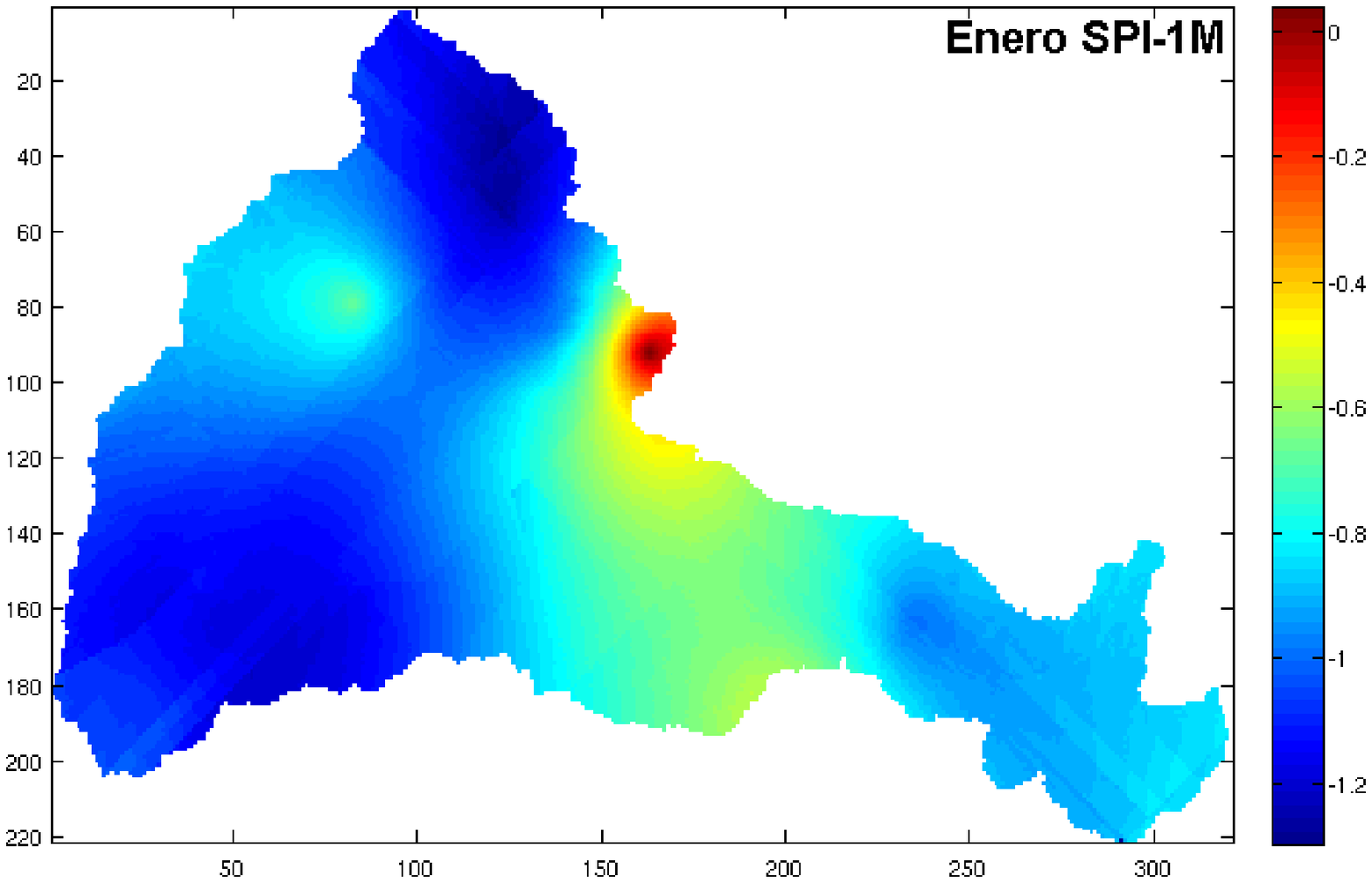}
\includegraphics[width=160pt]{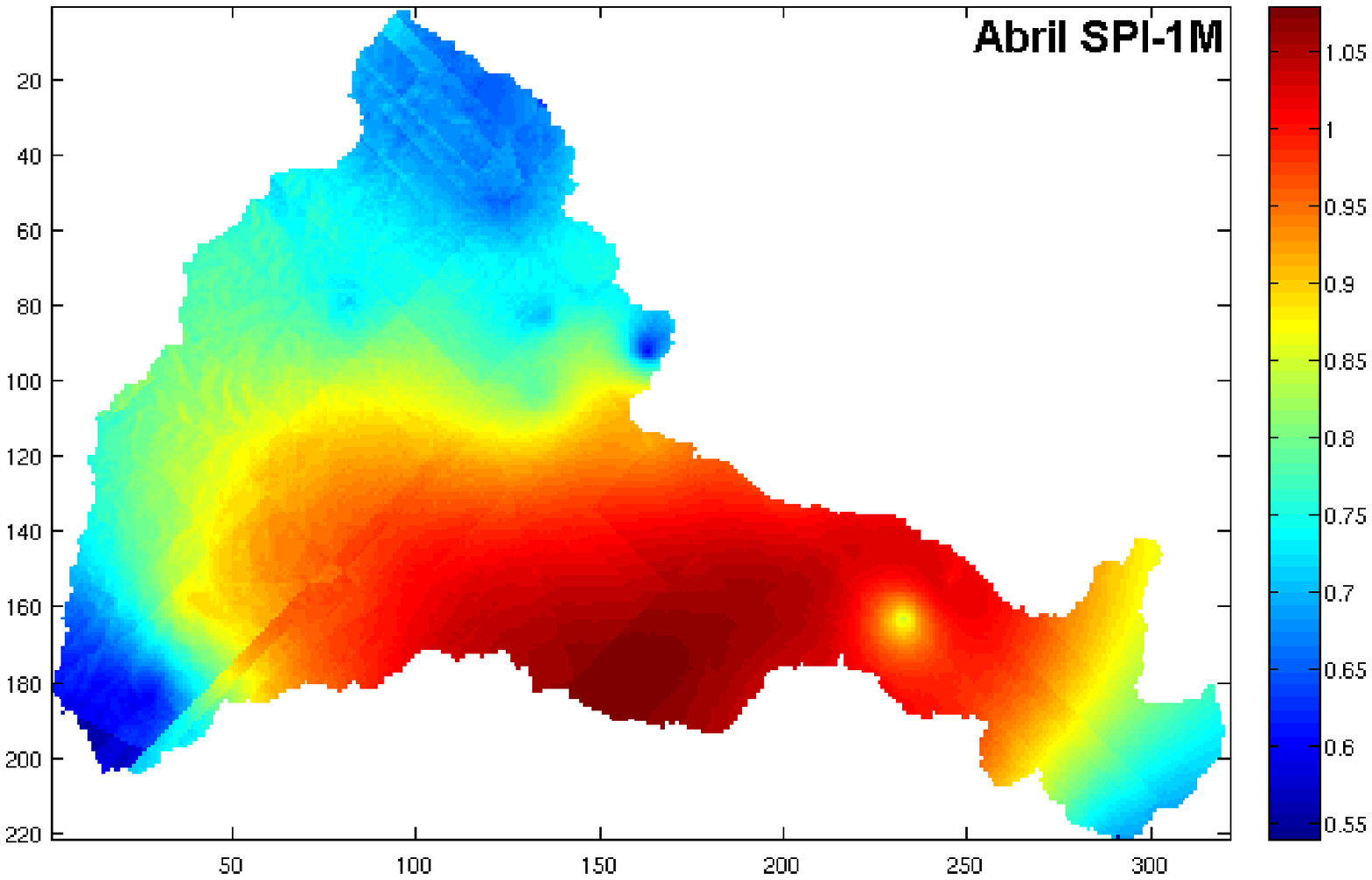}
\includegraphics[width=160pt]{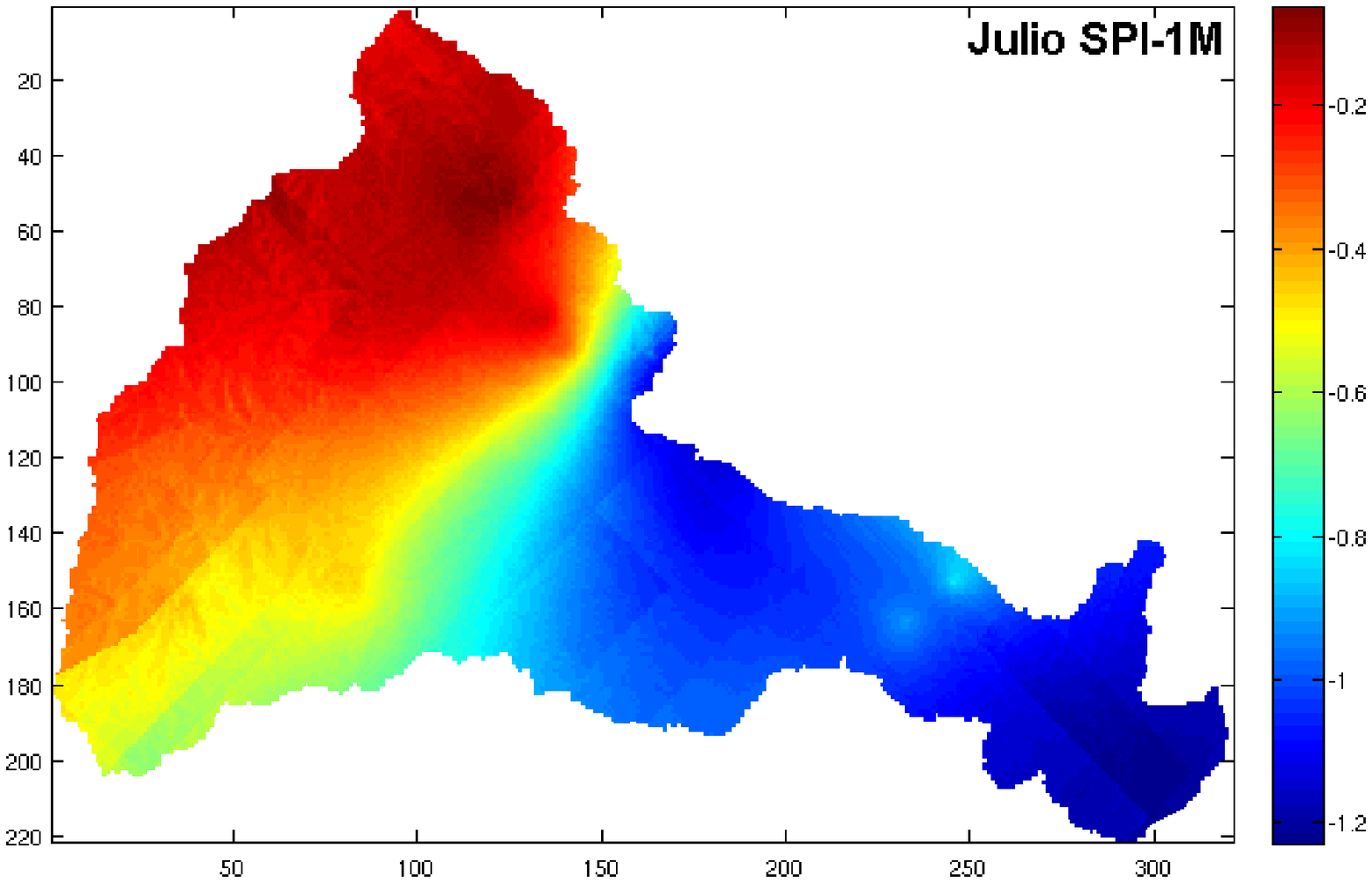}
\includegraphics[width=160pt]{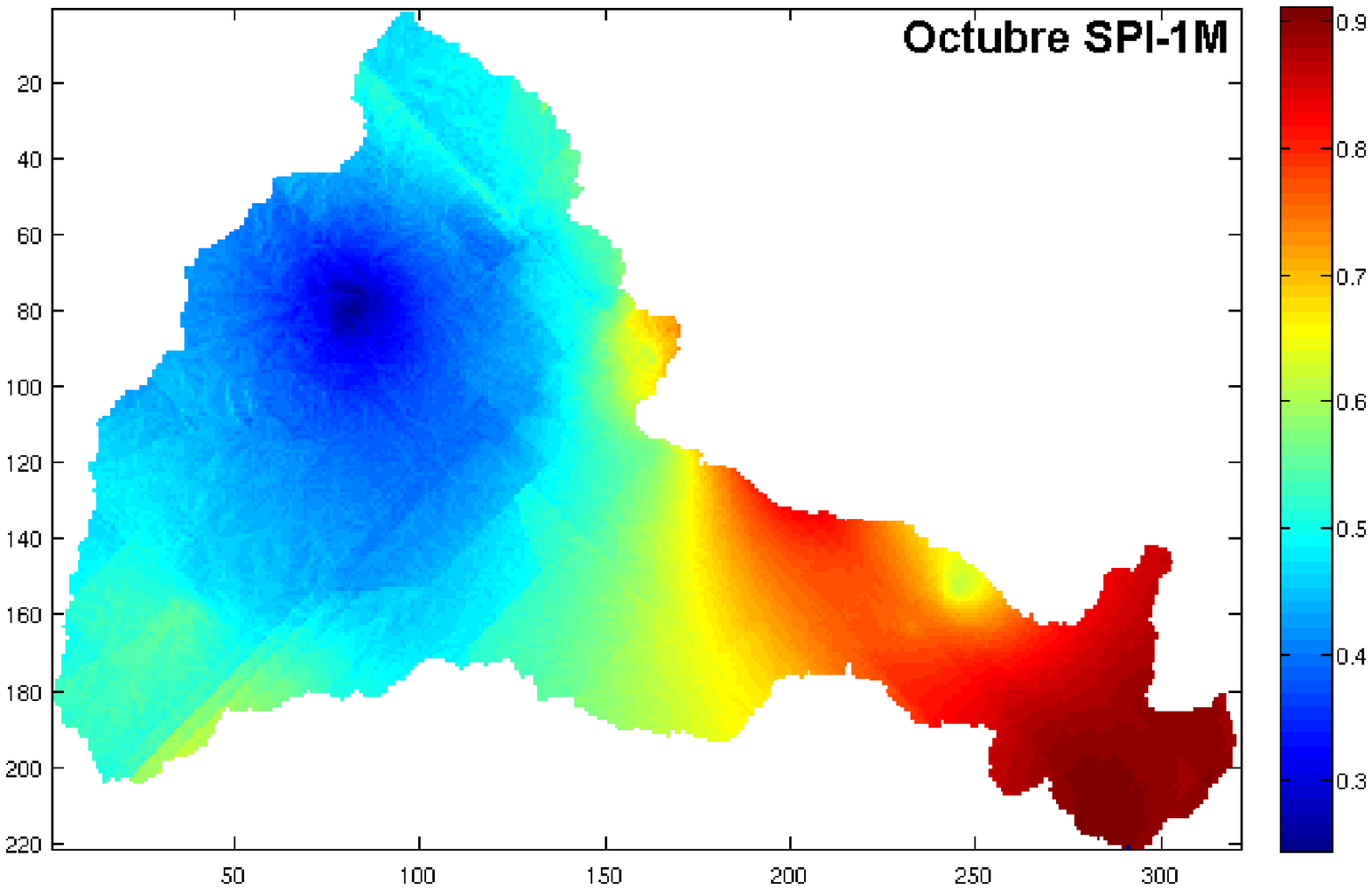}
\includegraphics[width=160pt]{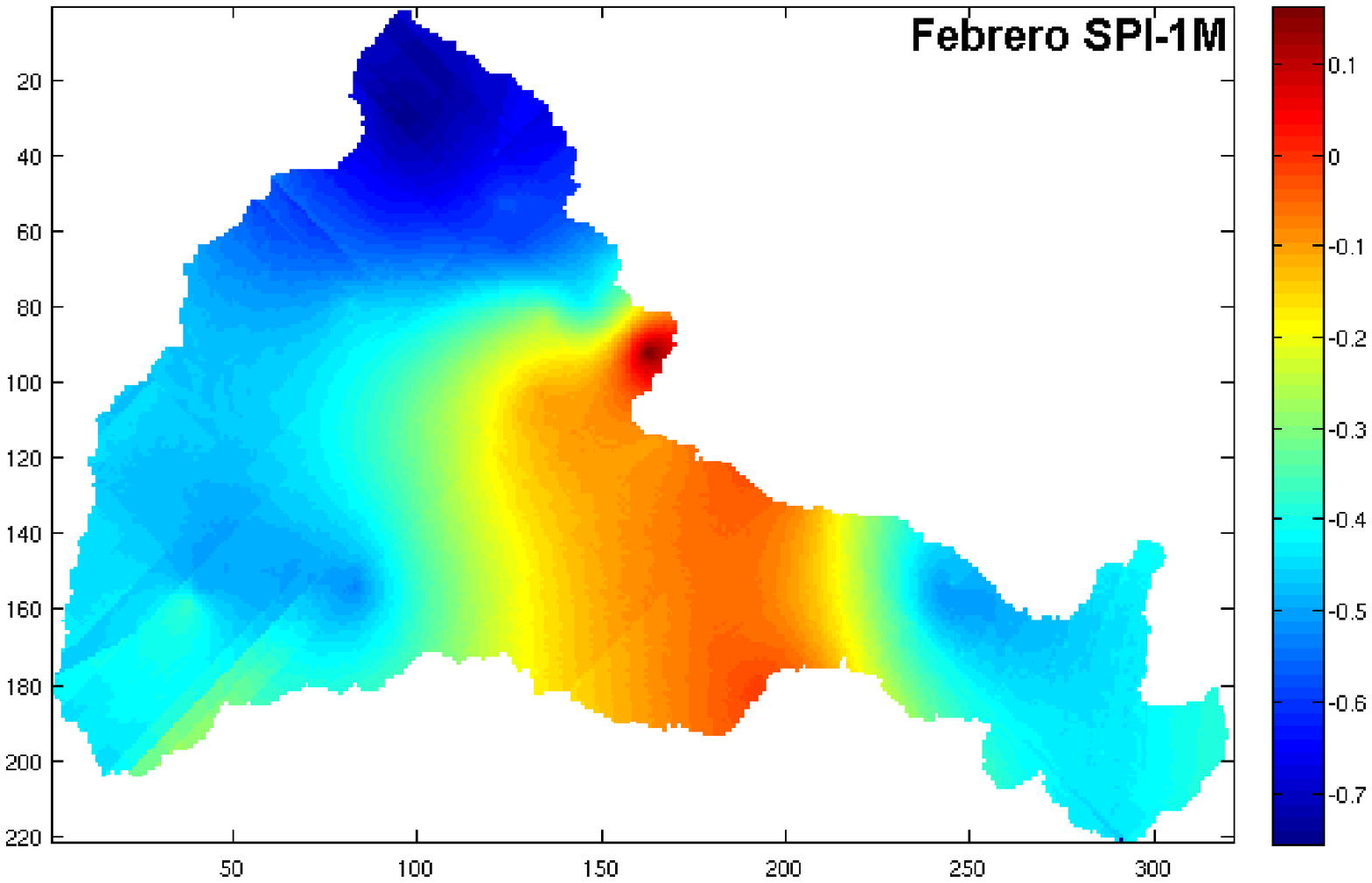}
\includegraphics[width=160pt]{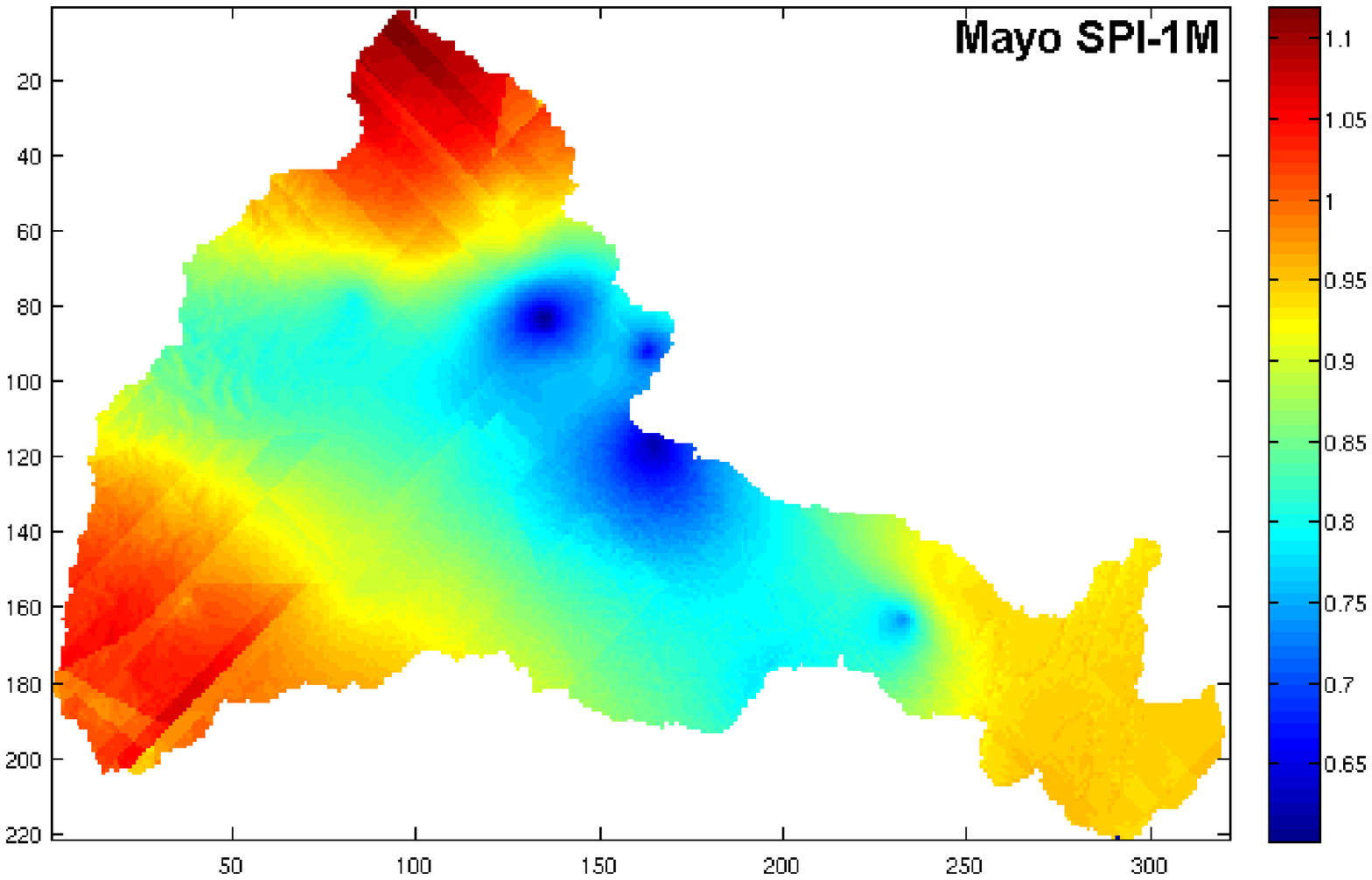}
\includegraphics[width=160pt]{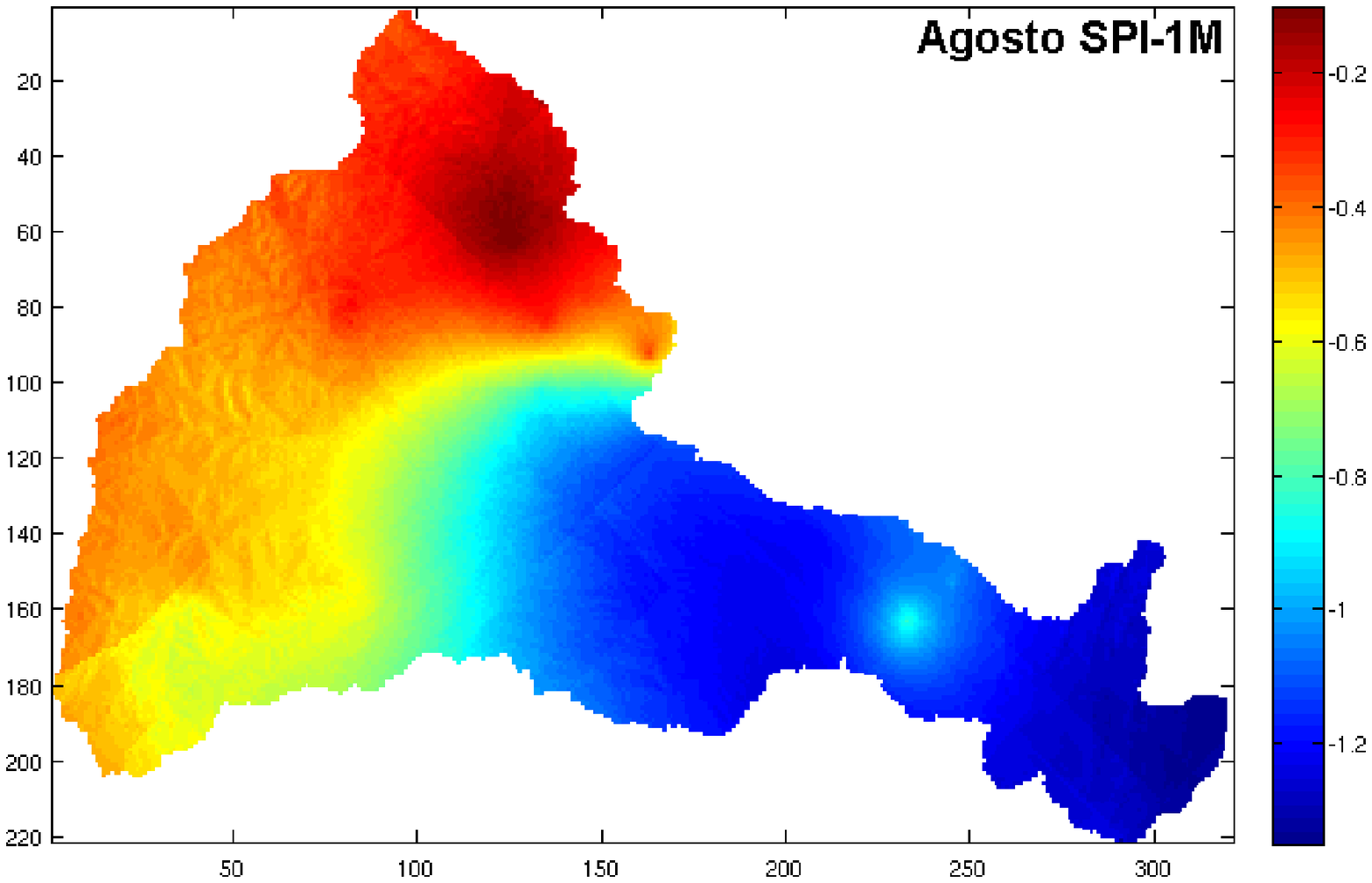}
\includegraphics[width=160pt]{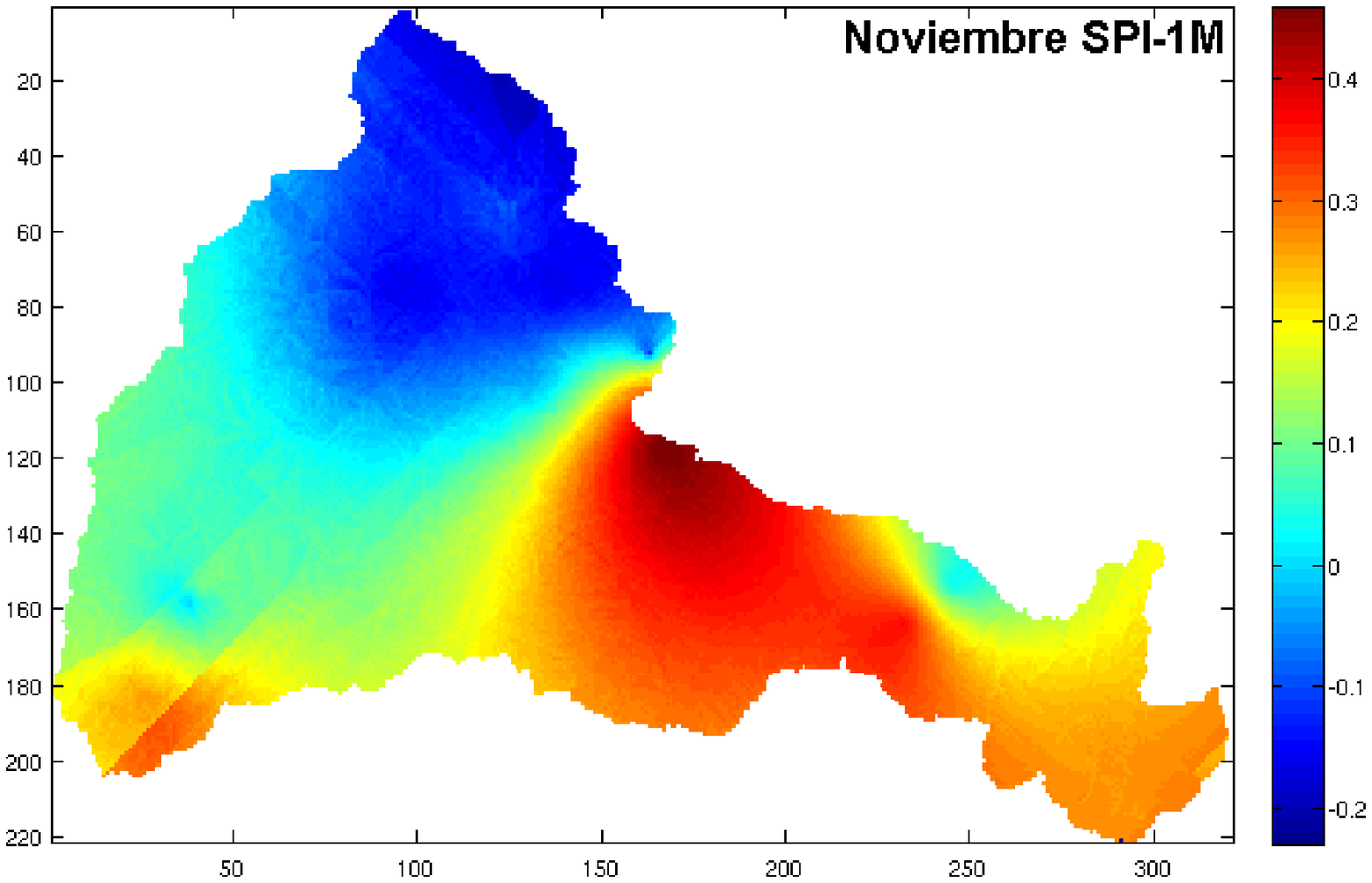}
\includegraphics[width=160pt]{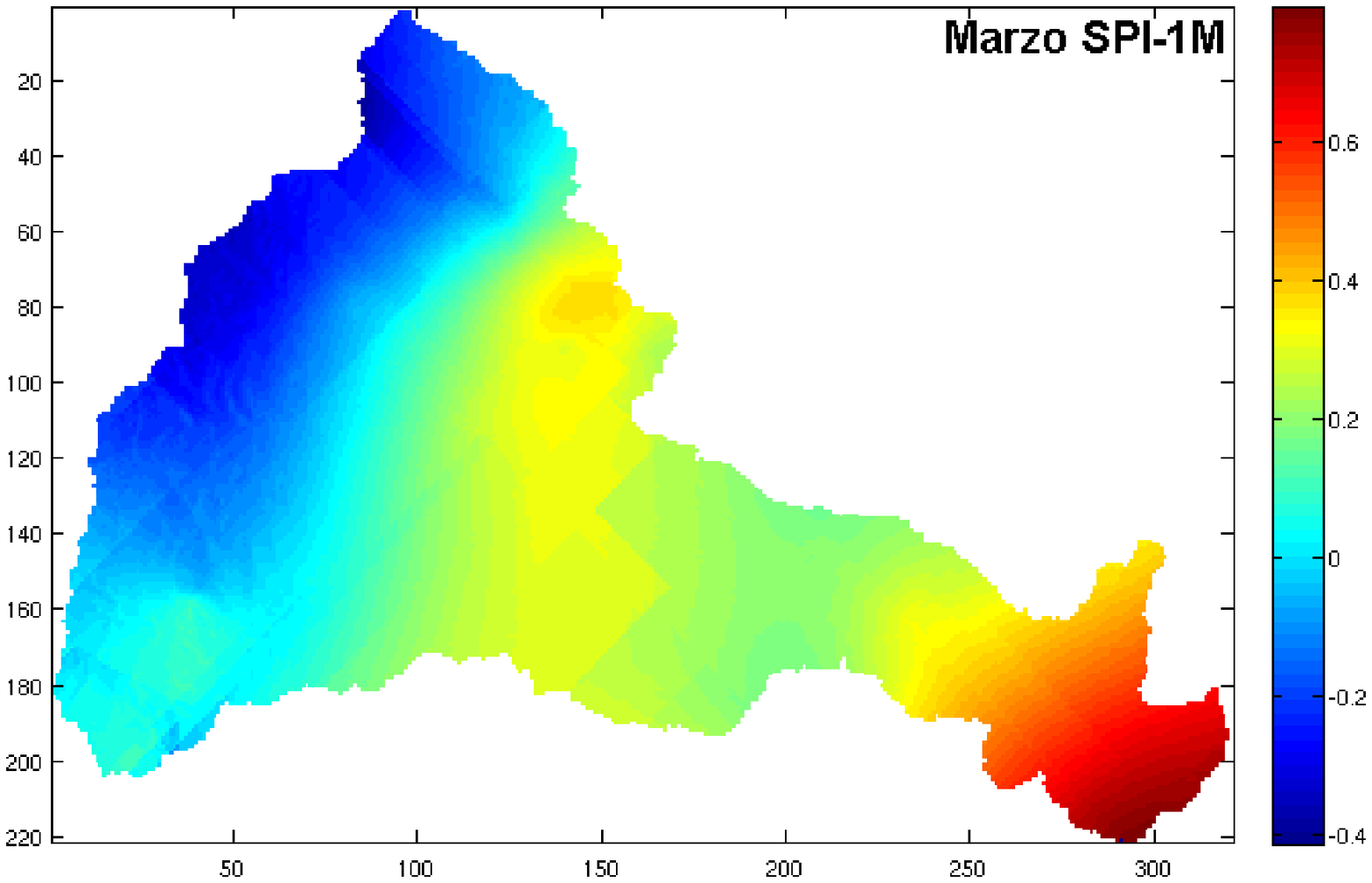}
\includegraphics[width=160pt]{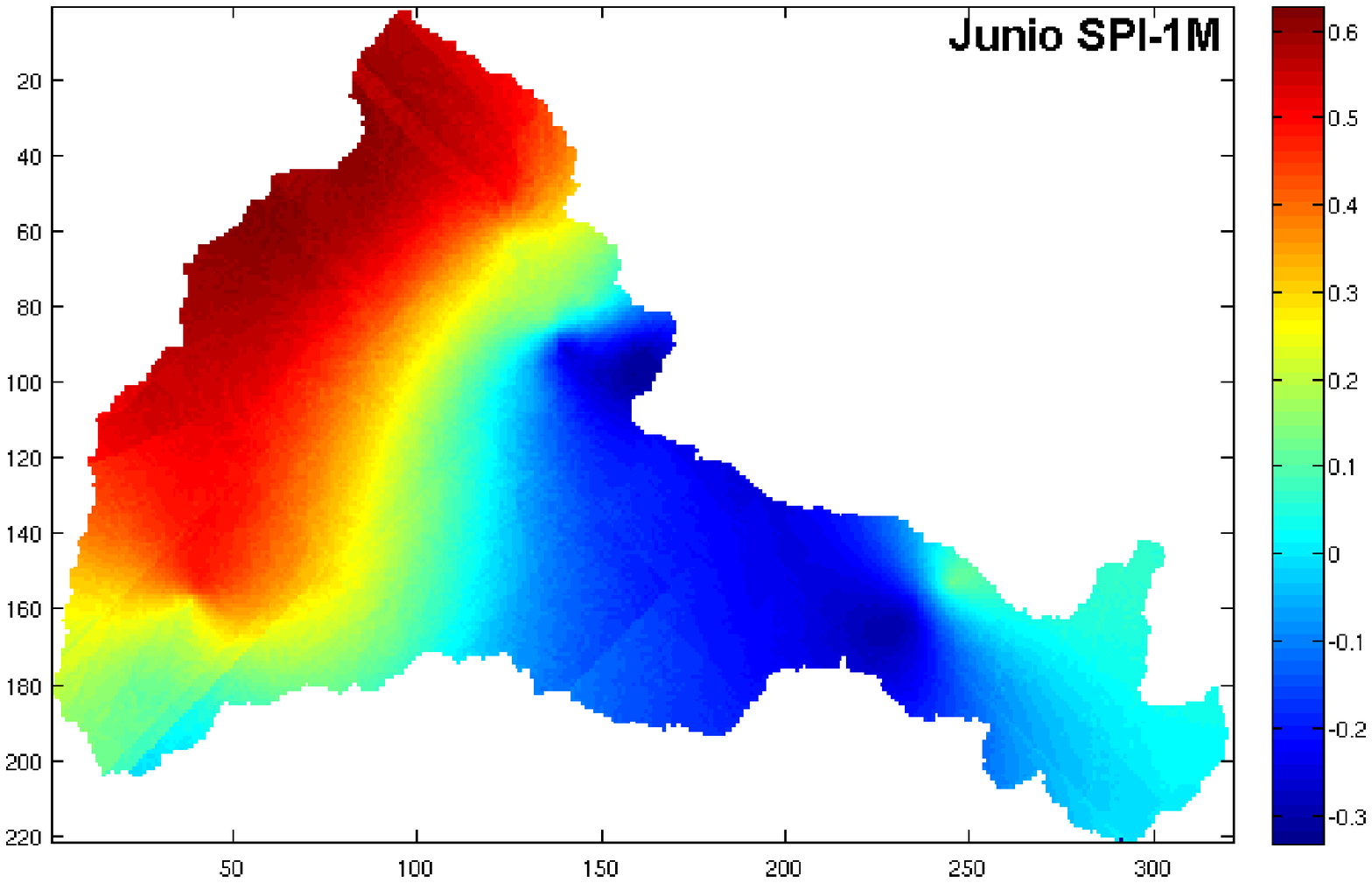}
\includegraphics[width=160pt]{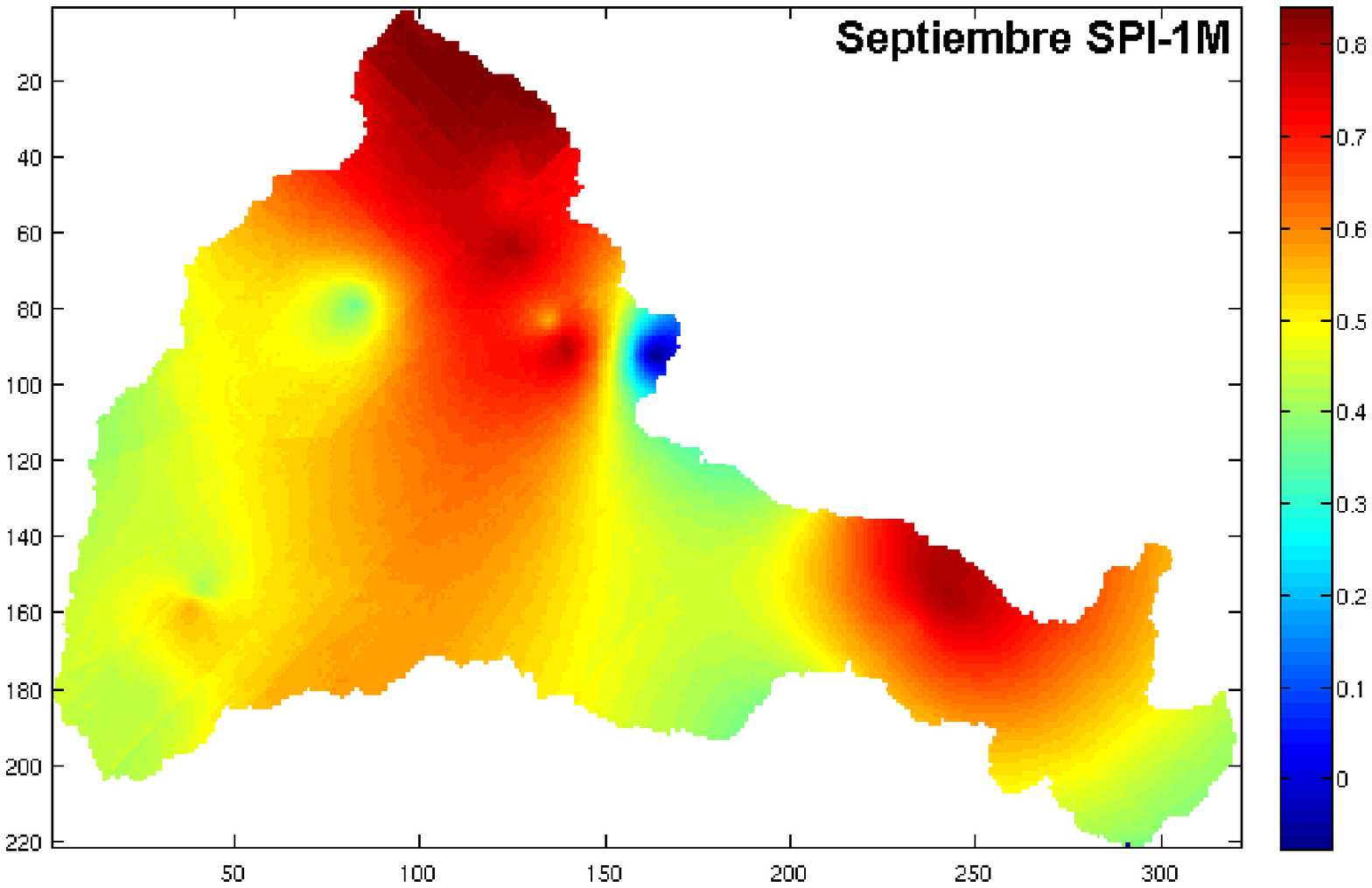}
\includegraphics[width=160pt]{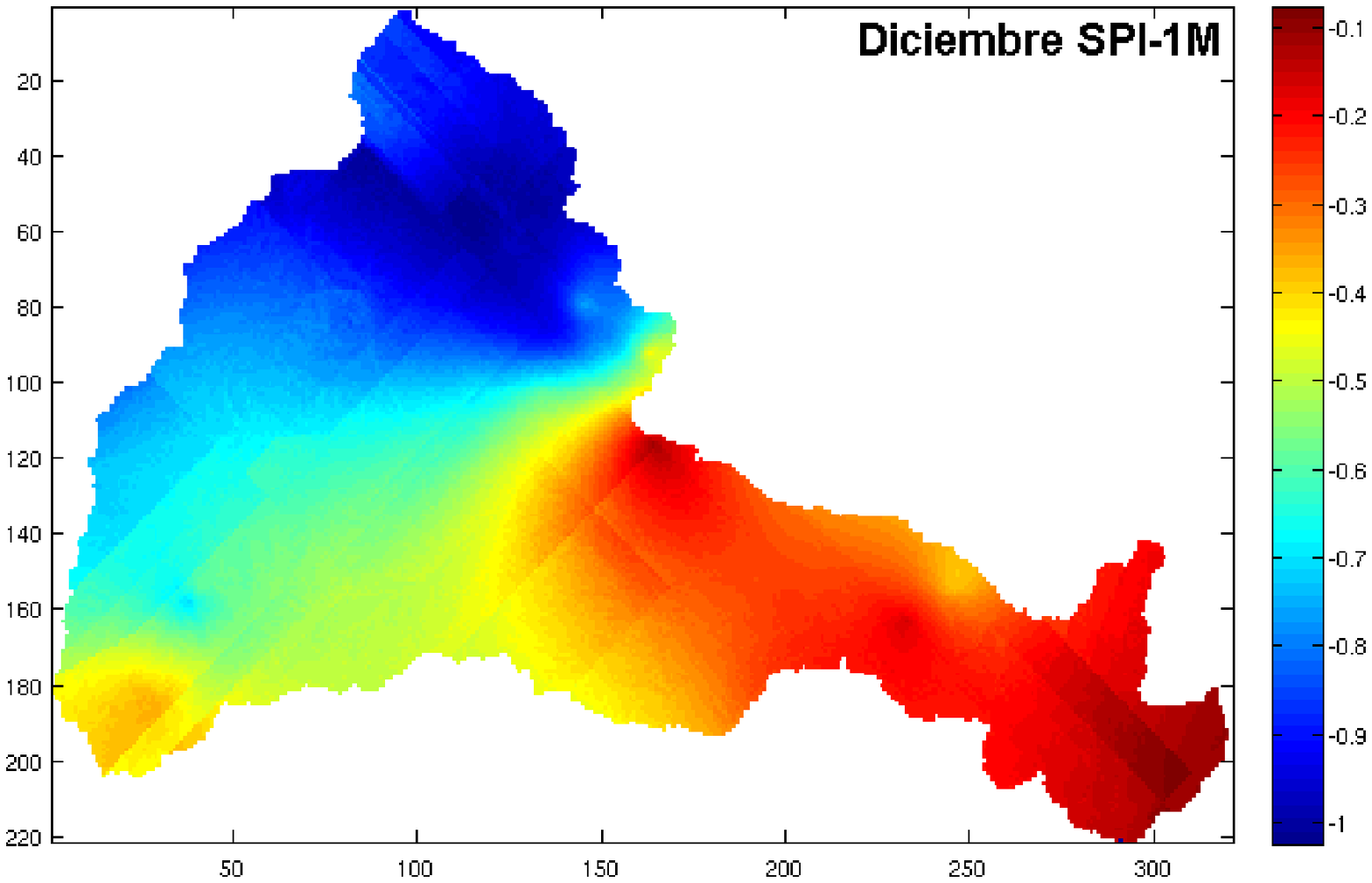}
\end{multicols}
\end{figure}

\newpage
\begin{center}
\textbf{APÉNDICE 2}
\end{center}
\bigskip
\bigskip
\bigskip
\bigskip

{Mapas de SPI a escala trimestral para la cuenca del río Coello.}

\begin{figure}[H]
 \begin{multicols}{3}
\includegraphics[width=160pt]{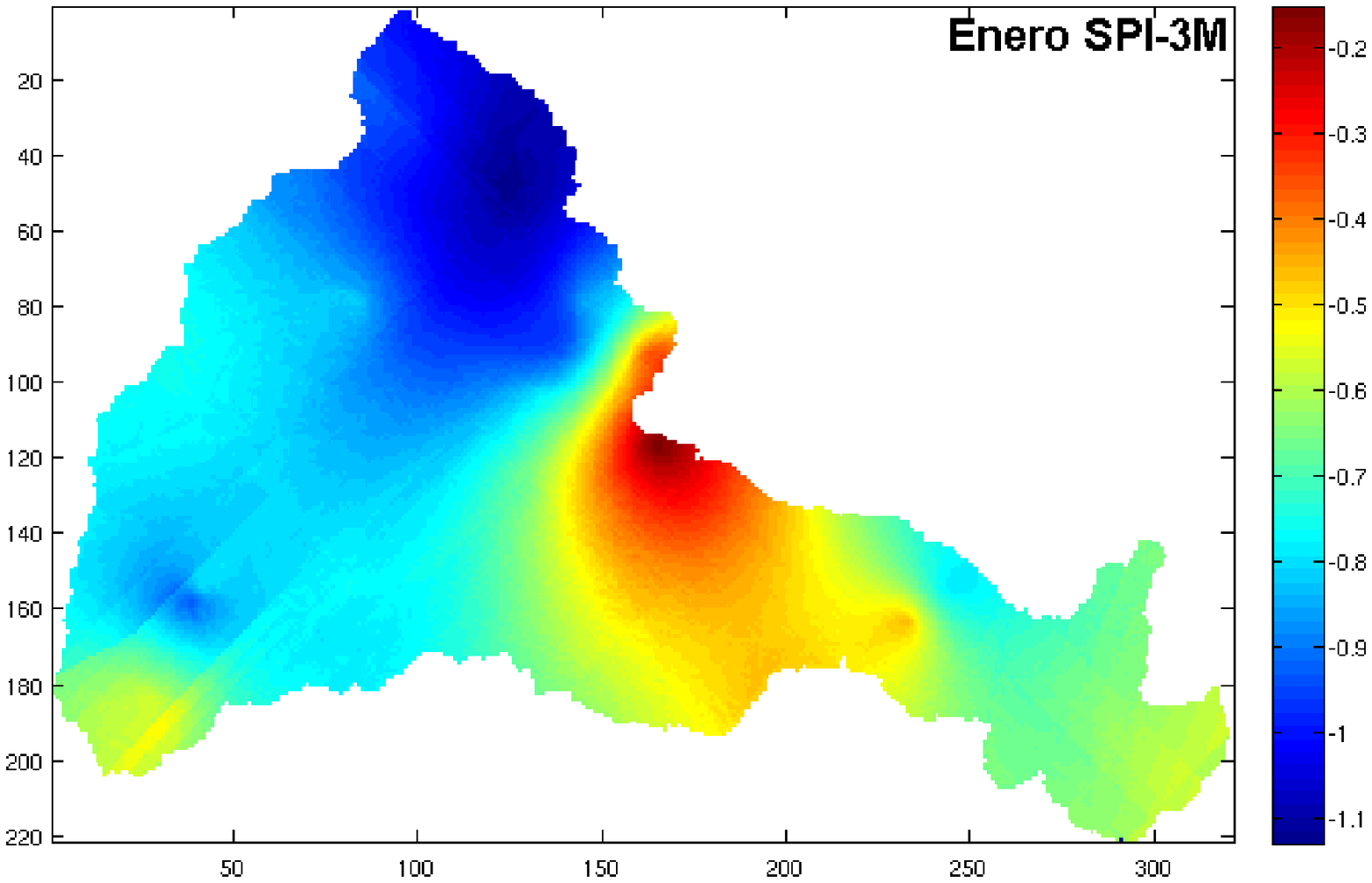}
\includegraphics[width=160pt]{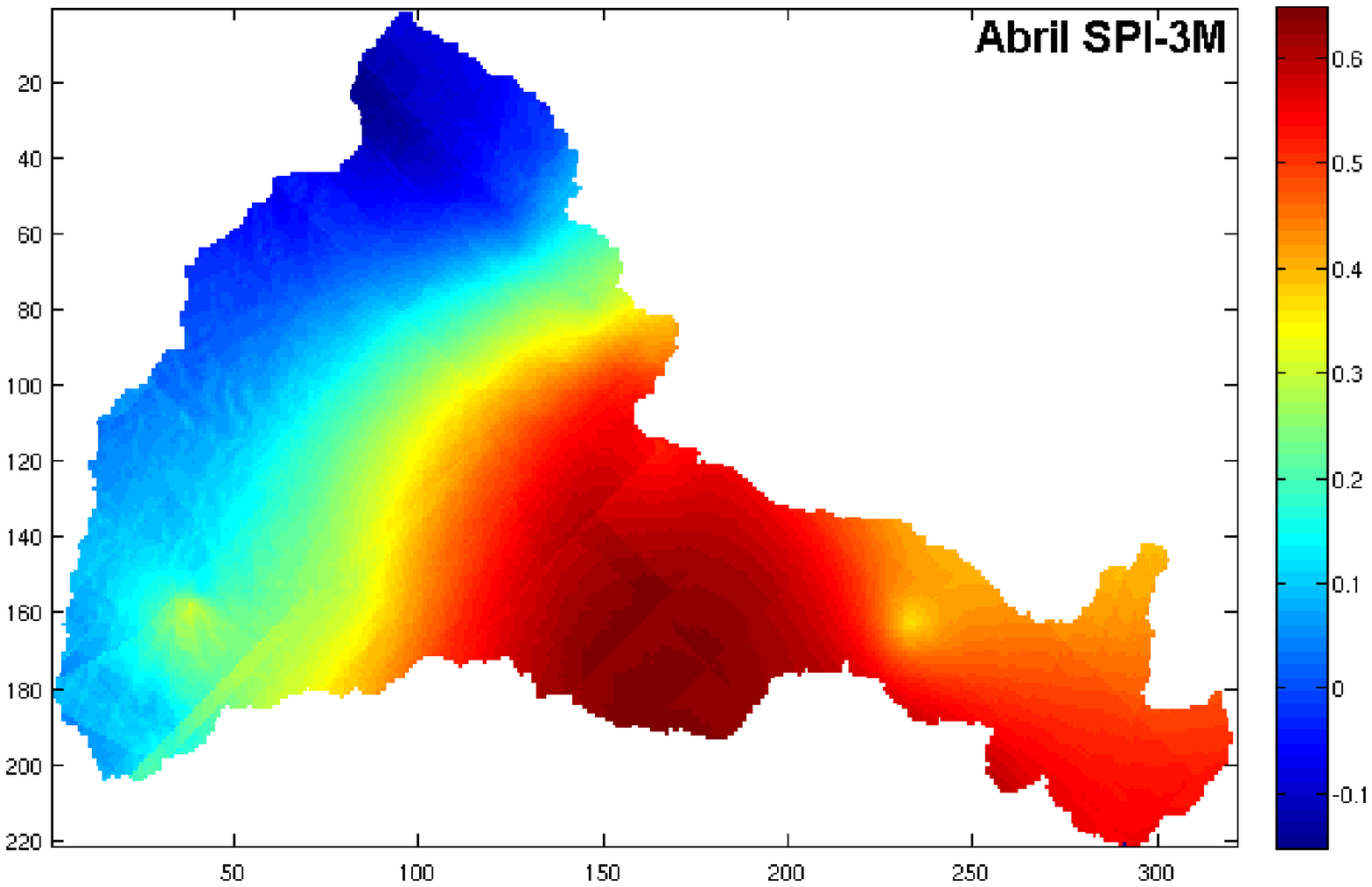}
\includegraphics[width=160pt]{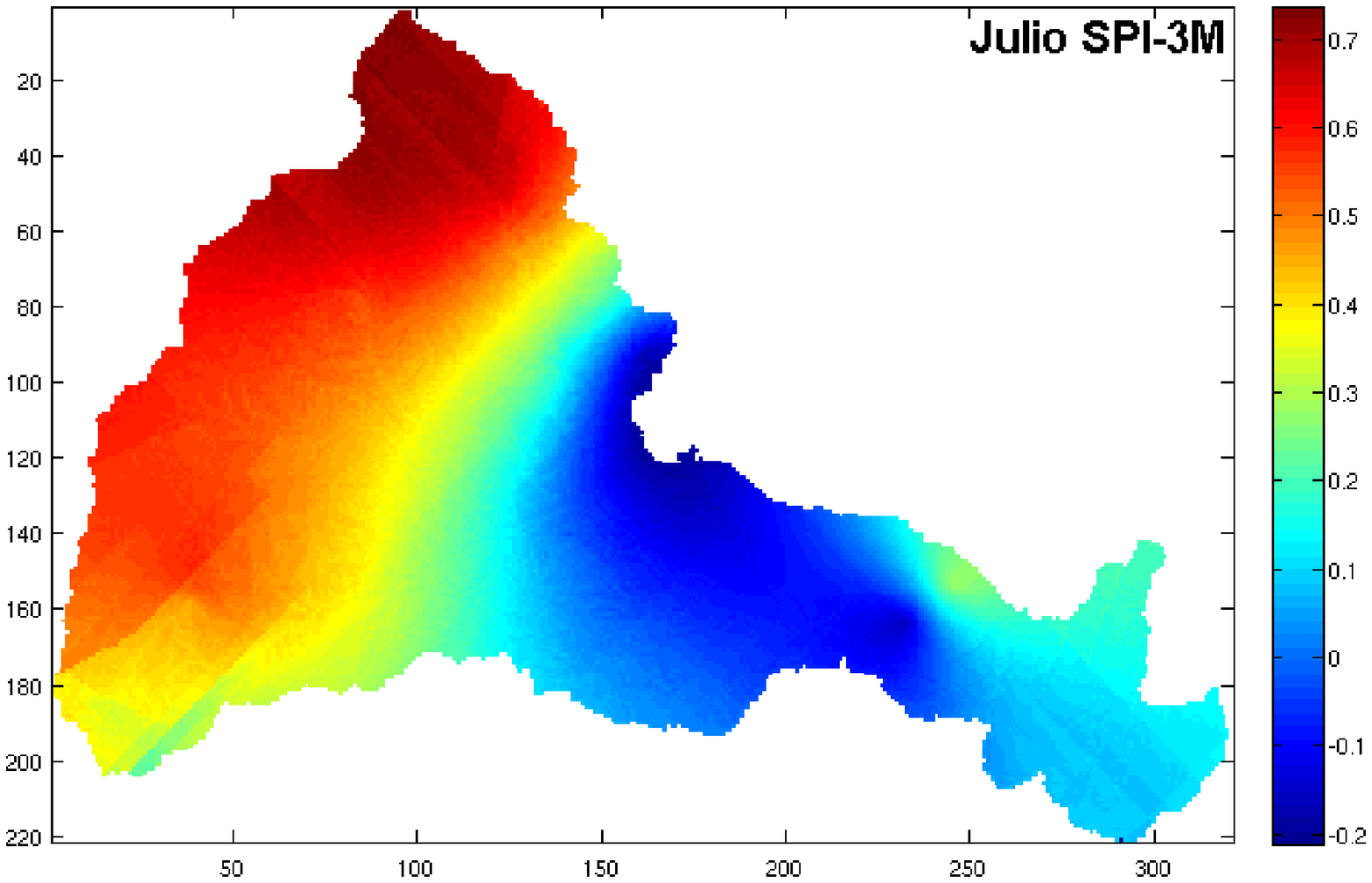}
\includegraphics[width=160pt]{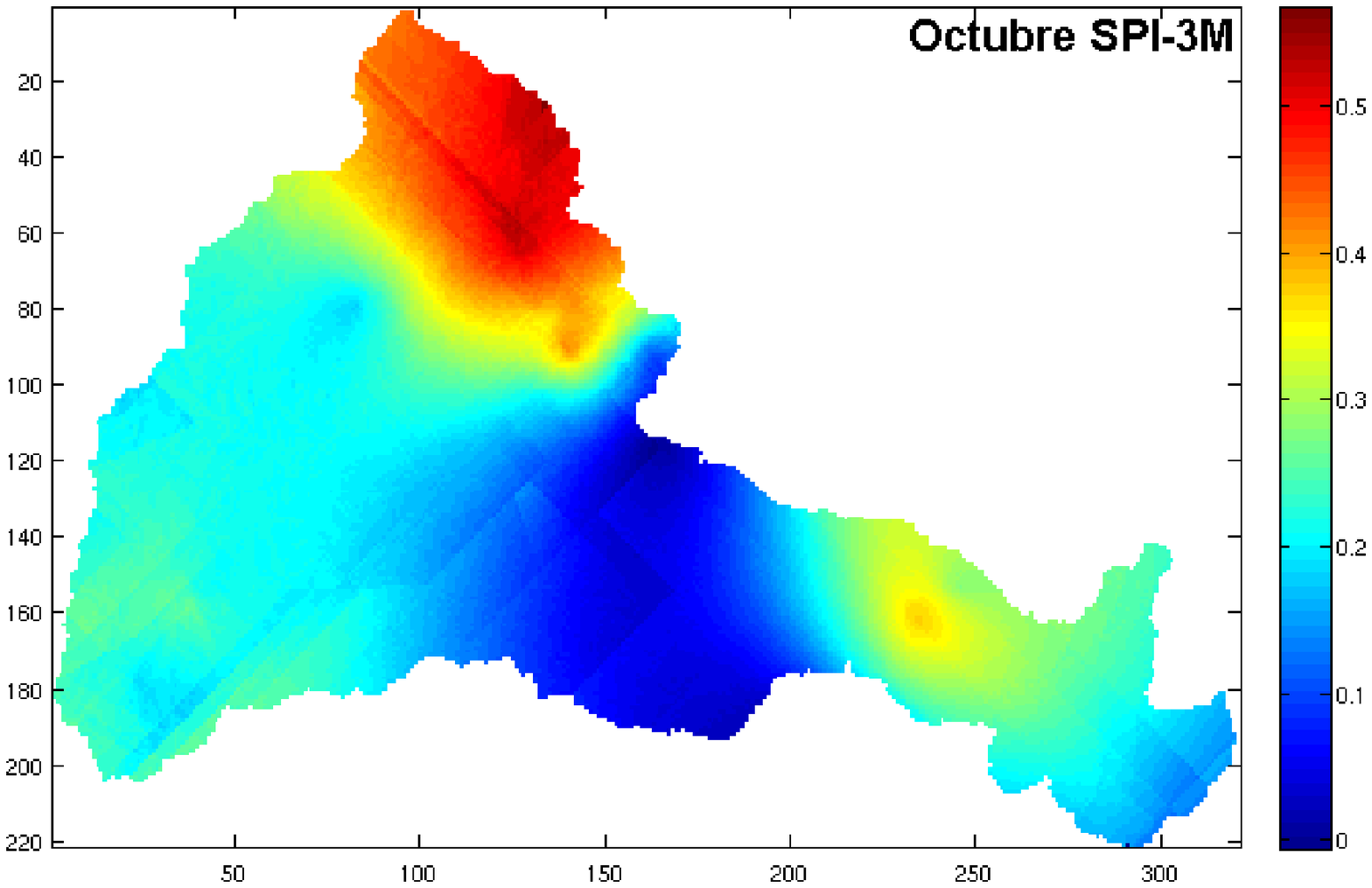}
\includegraphics[width=160pt]{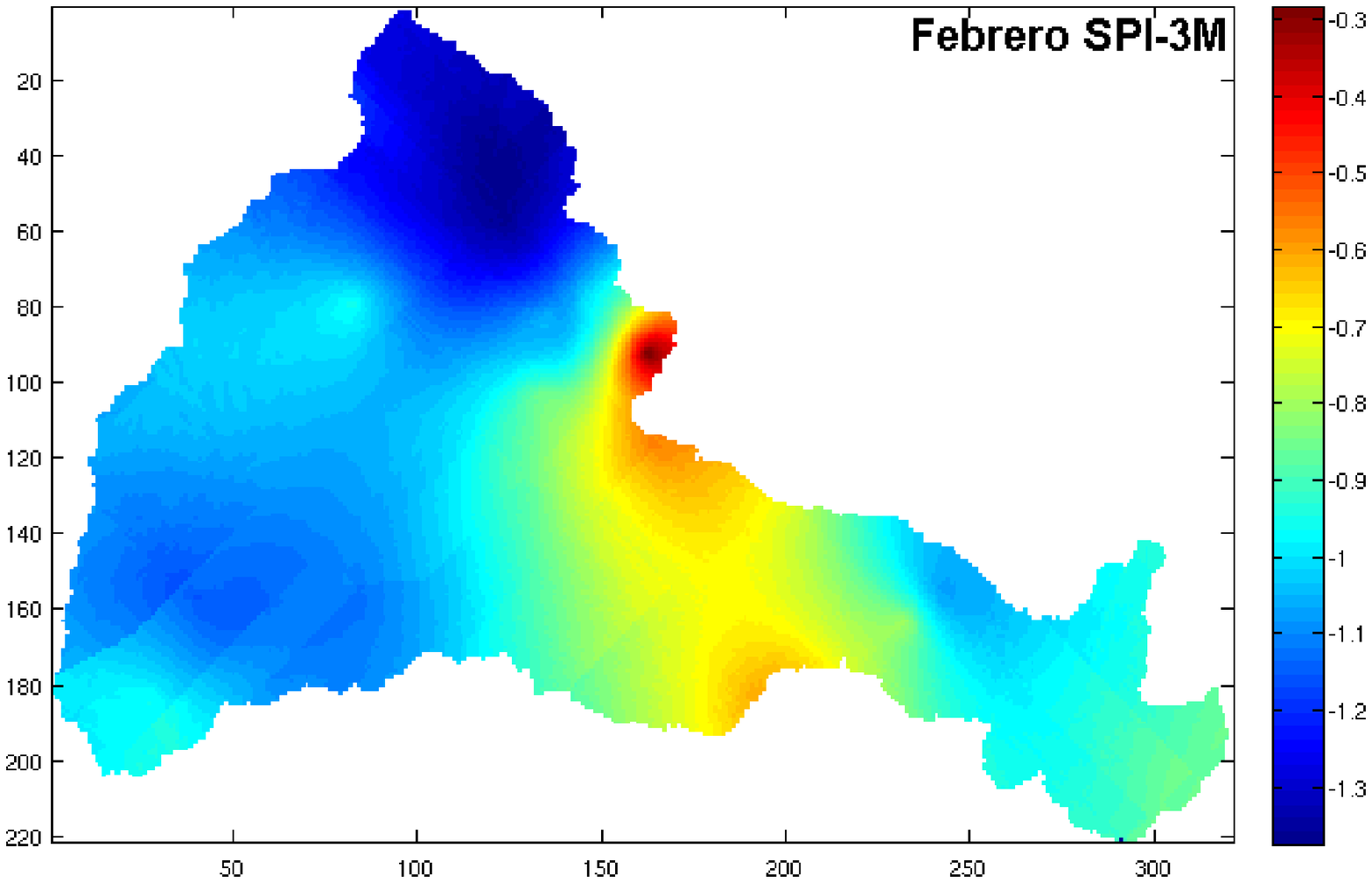}
\includegraphics[width=160pt]{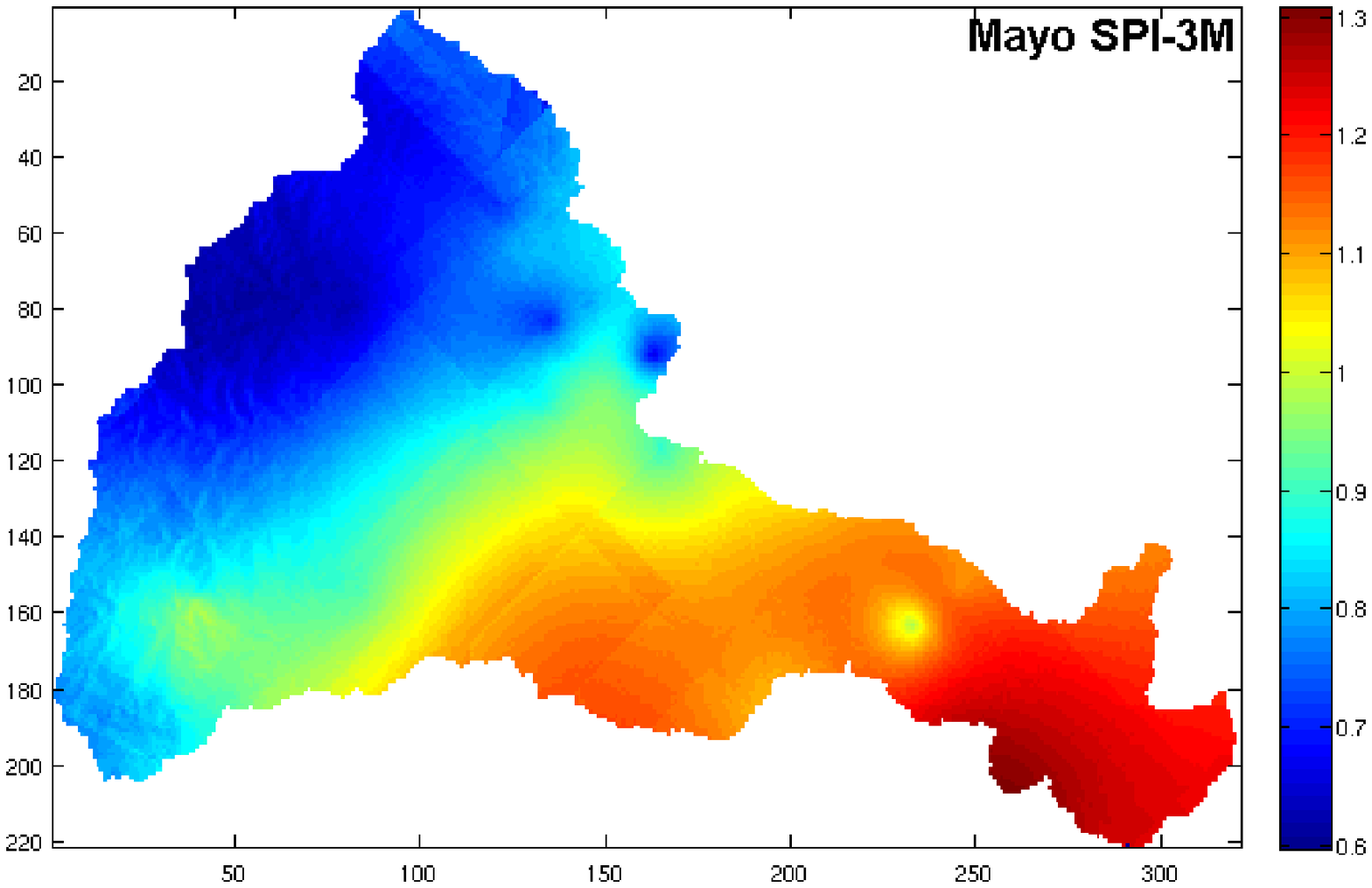}
\includegraphics[width=160pt]{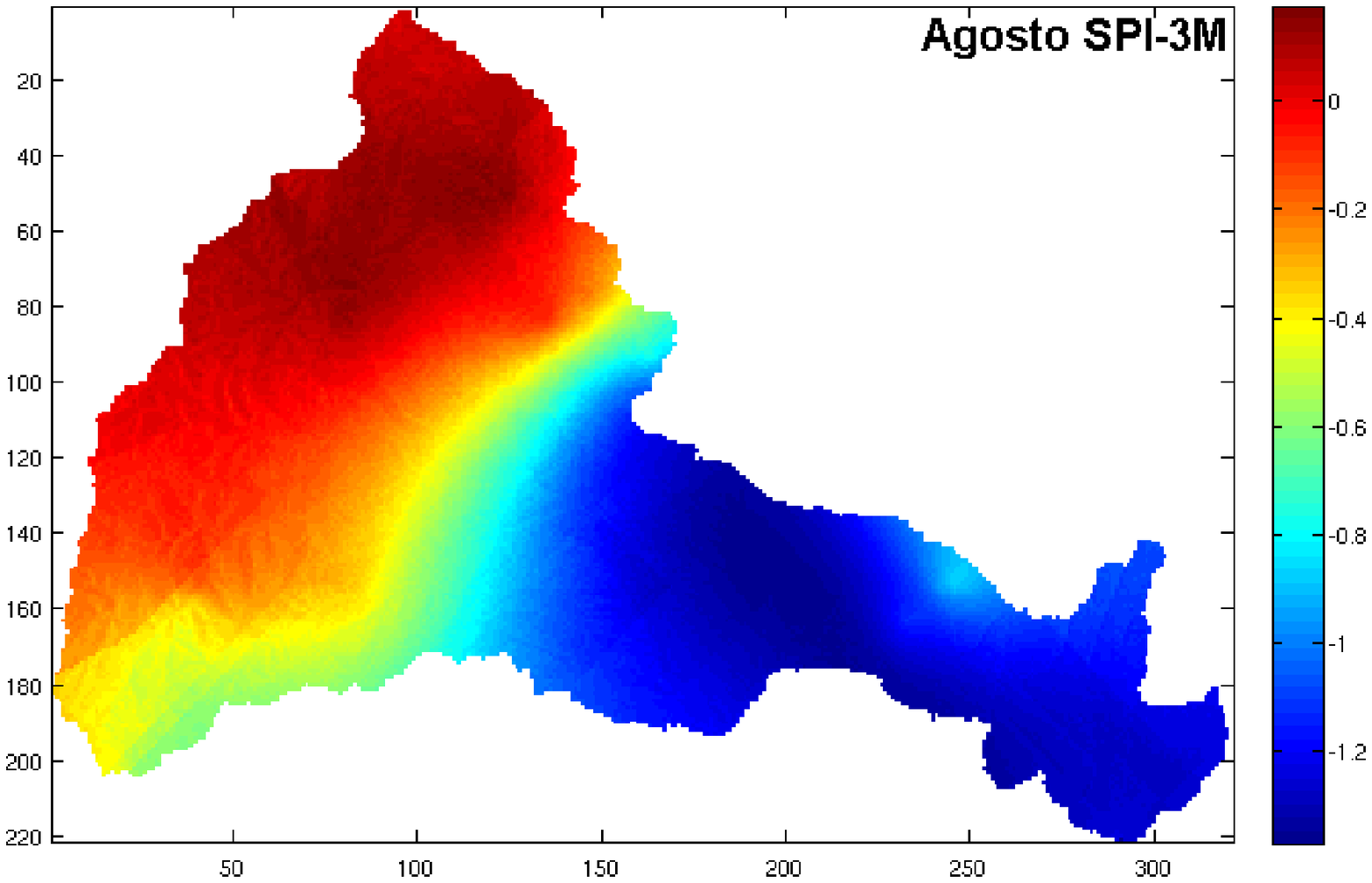}
\includegraphics[width=160pt]{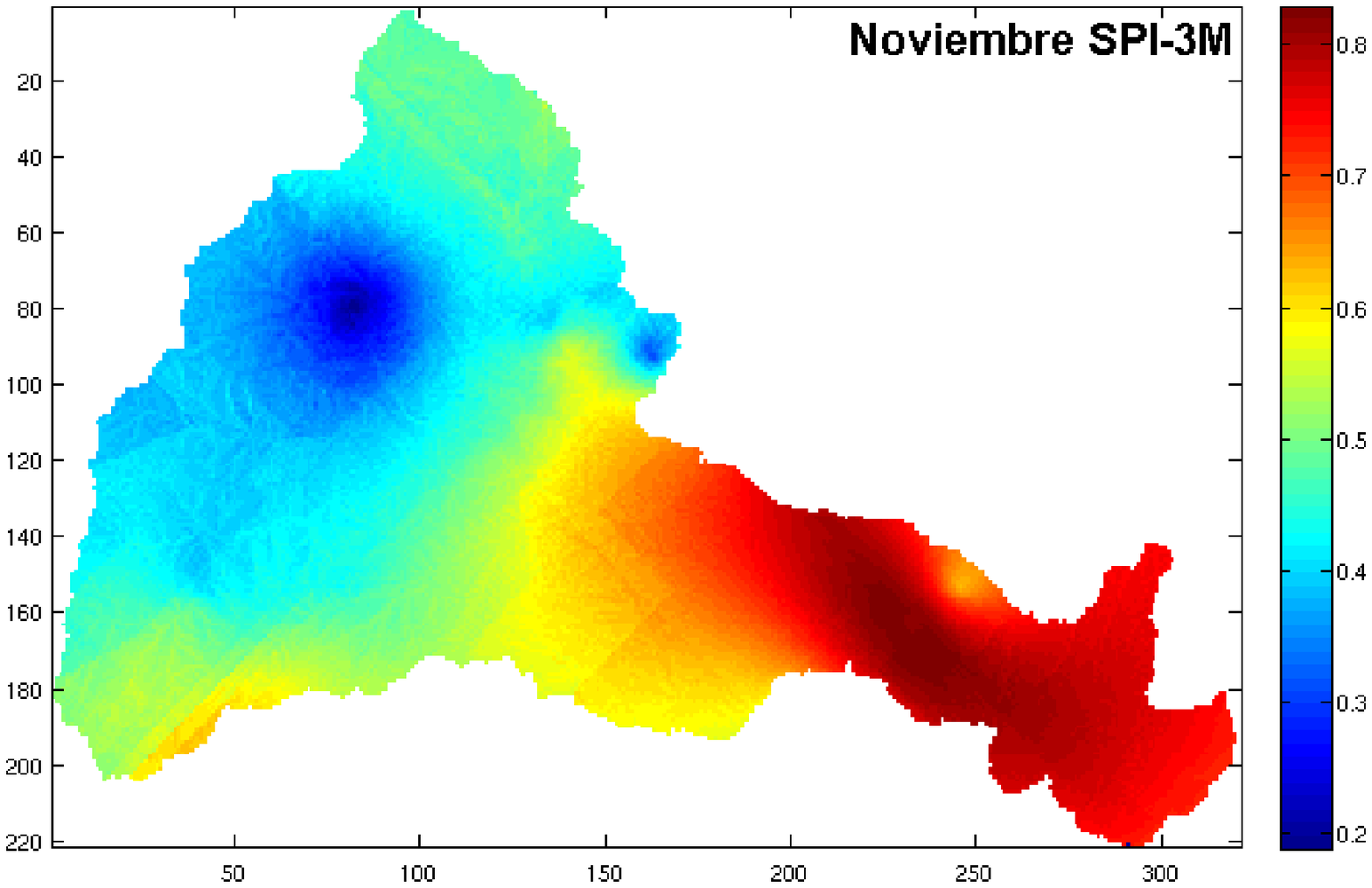}
\includegraphics[width=160pt]{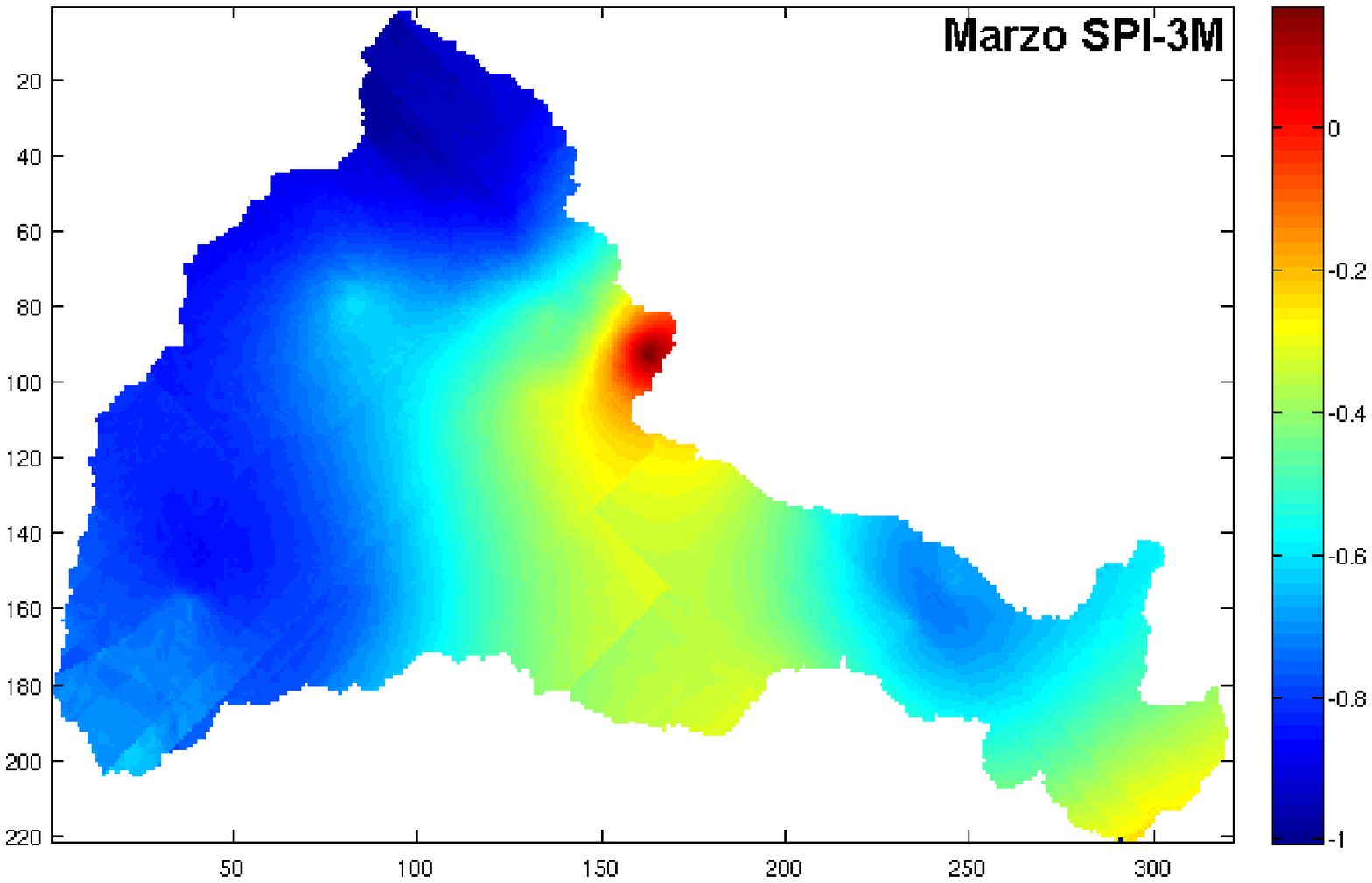}
\includegraphics[width=160pt]{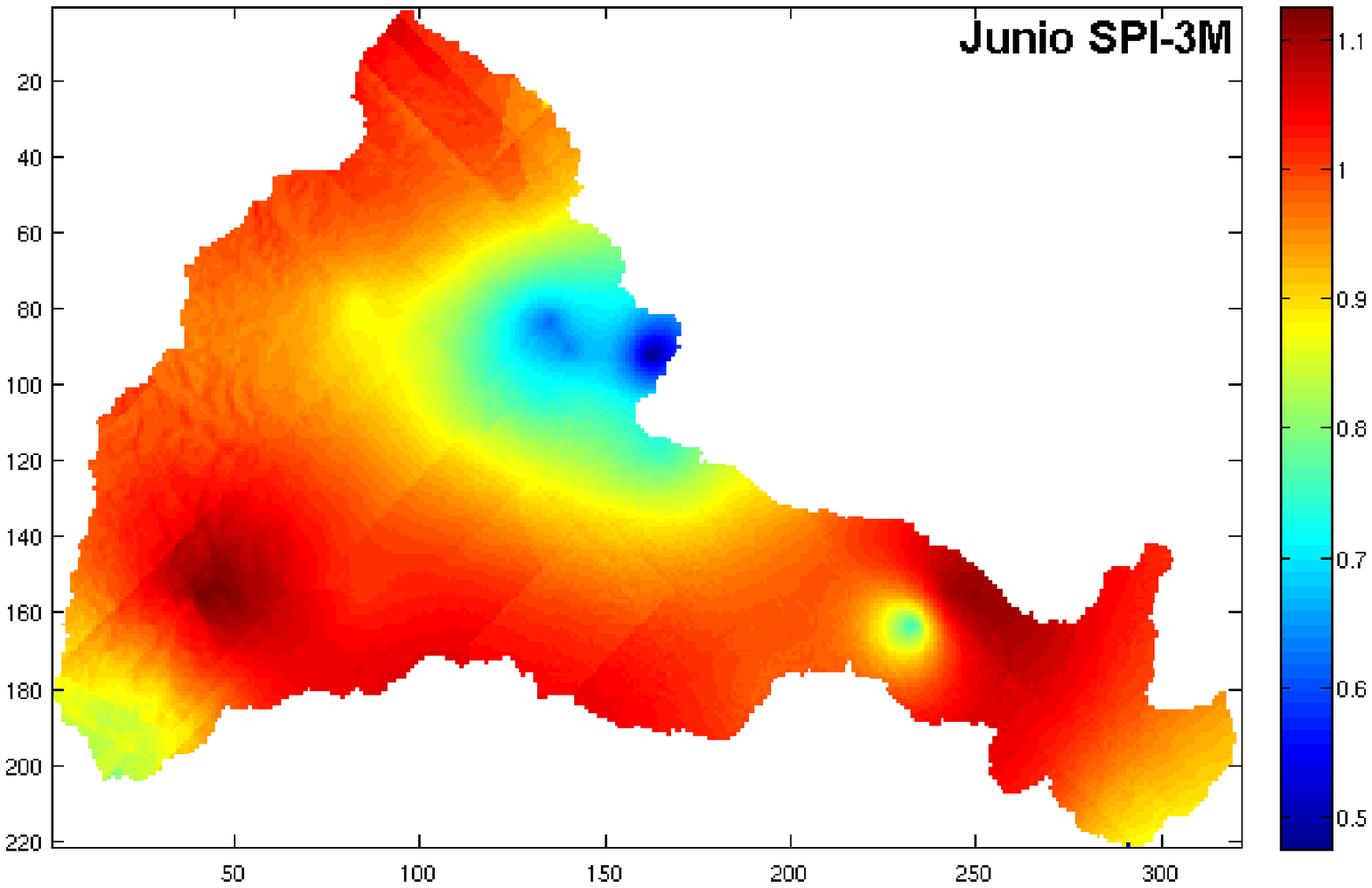}
\includegraphics[width=160pt]{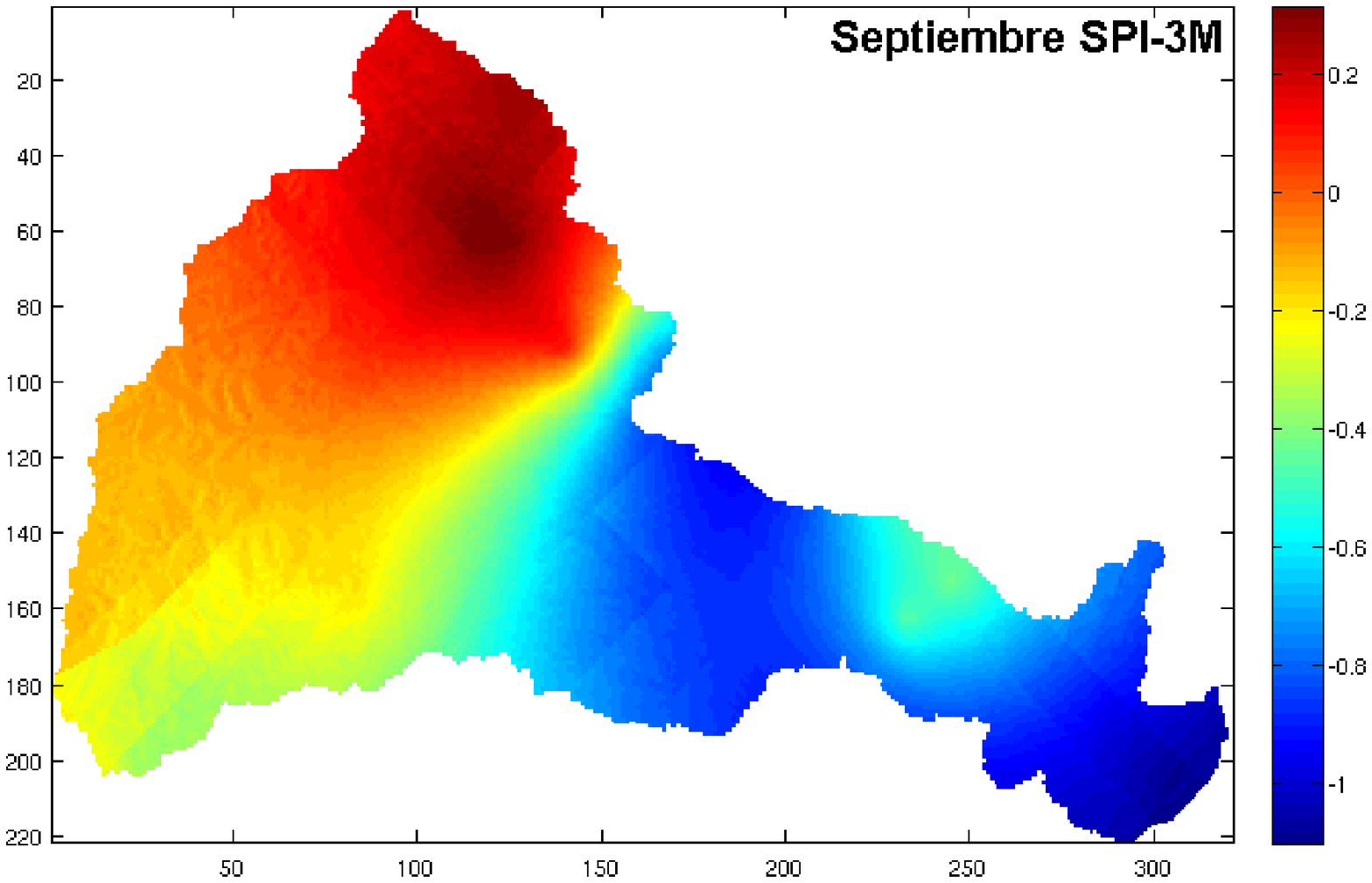}
\includegraphics[width=160pt]{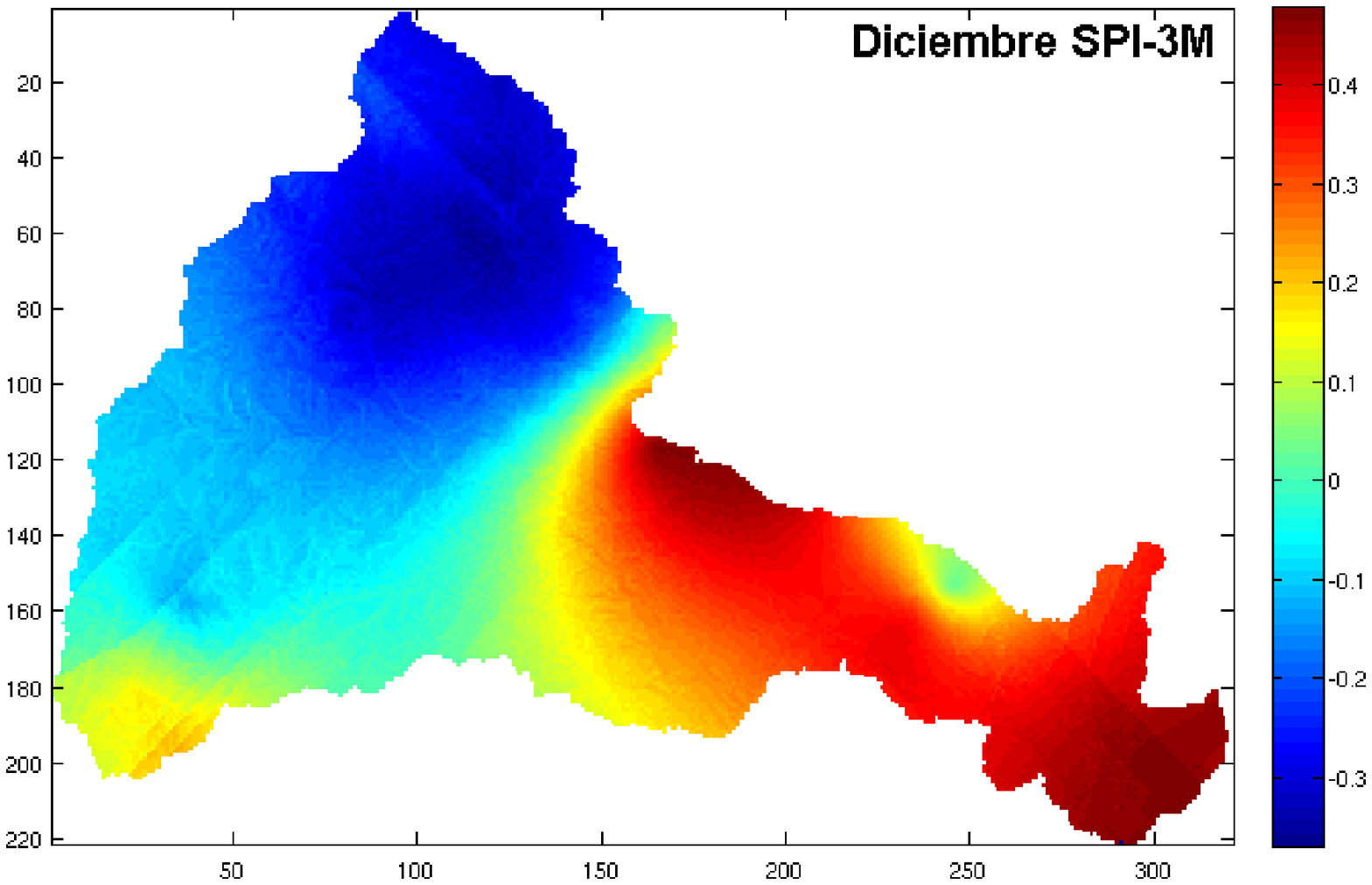}
\end{multicols}
\end{figure}

\newpage
\begin{center}
\textbf{APÉNDICE 3}
\end{center}
\bigskip
\bigskip
\bigskip
\bigskip

{Caracterización de sequías por medio de parámetros de SPI mensual y trimestral \textcolor{blue}{(Fig.10 a y b)}.}
\medskip

{La probabilidad acumulada expuesta en la columna 8 se refiere a la establecida por \textcolor{blue}{Mckee $et$ $al$ (1995)} para los intervalos de valores por el índice SPI.}

\begin{table}[H] \centering \scriptsize{
        \begin{tabular}{ccclcccc}
\toprule
\textbf{Índice} & \textbf{Categoría} & \textbf{N.Sequía} & \textbf{Duración} & \textbf{Intensidad} & \textbf{Severidad} & \textbf{Frecuencia}  & \textbf{Probabilidad}\\
\midrule

\multirow{29}{1cm}{\textbf{SPI-1M}}
&  & 1 & 50 dias & -0,95 & 1,03 & \\ 
&  & 2 & 1 mes & -0,18 & 0,18 & \\ 
&  & 6 & 45 dias & -0,73 & 0,83 &  \\
&  & 7 & 25 dias & -0,66 & 0,66 &  \\ 
&  & 9 & 15 dias & -0,57 & 0,57 & \\ 
&  & 10 & 40 dias & -0,37 & 0,37 & \\
& D1 & 11 & 2 meses & -0,93 & 1,76 & 13 en 10 años & 60\%\\ 
&  & 12 & 10 dias & -0,24 & 0,24 & \\ 
&  & 19 & 70 dias & -0,95 & 1,31 & \\
&  & 21 & 7 dias & -0,04 & 0,04 & \\
&  & 22 & 15 dias & -0,34 & 0,34 & \\
&  & 23 & 1 mes & -0,39 & 0,39 & \\
&  & 26 & 1 mes & -0,6 & 0,67 & \\ \cmidrule{2-8}
&  & 8 & 1 mes & -1,39 & 1,39 & \\
&  & 14 & 4 meses & -1,16 & 2,68 & \\
&  & 15 & 35 dias & -1,03 & 1,03 & \\
& D2 & 18 & 45 dias & -1,32 & 1,33 & 7 en 10 años & 40\%\\
&  & 25 & 1 mes & -1,43 & 1,43 & \\
&  & 28 & 2 meses & -1,01 & 1,71 & \\
&  & 29 & 45 dias & -1,32 & 1,32 & \\\cmidrule{2-8}
&  & 4 & 4 meses & -1,59 & 3,5 & \\
&  & 16 & 45 dias & -1,57 & 1,57 & \\
& D3 & 17 & 4 meses & -1,58 & 5,15 & 6 en 10 años & 20\%\\
&  & 20 & 4 meses & -1,52 & 2,72 & \\
&  & 24 & 3 meses & -1,68 & 1,97 & \\
&  & 27 & 75 dias & -1,56 & 2,14 & \\ \cmidrule{2-8}
&  & 3 & 4 meses & -2,55 & 4,99 & \\ 
& D4 & 5 & 3 meses & -2,3 & 5,53 & 3 en 10 años & 10\%\\
&  & 13 & 2 meses & -2,01 & 2,24 & \\ \cmidrule{1-8}
\multirow{16}{1cm}{\textbf{SPI-3M}}
&  & 1 & 2 meses & -0,12 & 0,12 & \\ 
&  & 2 & 2 meses & -0,07 & 0,11 & \\
&  & 4 & 2 meses & -0,43 & 0,45 & \\
&  & 5 & 1 meses & -0,36 & 0,36 & \\
& D1 & 7 & 45 dias & -0,68 & 0,68 & 9 en 10 años & 60\%\\ 
&  & 9 & 5 meses & -0,55 & 1,61 & \\
&  & 11 & 45 dias & -0,32 & 0,32 & \\
&  & 12 & 6 meses & -0,86 & 2,59 &  \\
&  & 13 & 1 mes & -0,34 & 0,34 & \\ \cmidrule{2-8}
&  & 6 & 8 meses & -1,23 & 5,55 & \\
& D2 & 10 & 3 meses & -1,06 & 2,8 & 3 en 10 años & 40\%\\
&  & 14 & 6 meses & -1,31 & 5,36 & \\ \cmidrule{2-8}
& D3 & 8 & 7 meses & -1,69 & 5,85 & 1 en 10 años & 20\%\\ \cmidrule{2-8}
& D4 & 3 & 14 meses & -2,57 & 20,43 & 1 en 10 años & 10\%\\

\bottomrule
\end{tabular}} \end{table}

\newpage
\listoffigures
\listoftables

\bigskip
\bigskip
\bigskip
\bigskip
\bigskip
\bigskip\
\bigskip
\bigskip
Este artículo fue editado y redactado en el software \LaTeX .

\end{document}